\documentclass[a4paper,fleqn,usenatbib]{mnras}

\usepackage{newtxtext,newtxmath}
\usepackage{times}
\usepackage[T1]{fontenc}
\usepackage{ae,aecompl}
\usepackage{supertabular}
\usepackage{longtable}
\usepackage{amssymb}

\usepackage[dvipsnames]{xcolor}

\usepackage{caption}

\usepackage{etoolbox}
\makeatletter
\patchcmd\@combinedblfloats{\box\@outputbox}{\unvbox\@outputbox}{}{%
	\errmessage{\noexpand\@combinedblfloats could not be patched}%
}%
\makeatother

\usepackage{graphicx}	
\usepackage{amsmath}	
\usepackage{amssymb}
\usepackage{float}
\newcommand{\overbar}[1]{\mkern 1.5mu\overline{\mkern-1.5mu#1\mkern-1.5mu}\mkern 1.5mu}

\title[From birth associations to field stars]{From birth associations to field stars: mapping the small-scale orbit distribution in the Galactic disc}
\author[J. Coronado et al.]{Johanna Coronado,$^{1}$\thanks{E-mail: coronado@mpia.de} Hans-Walter Rix,$^{1}$  Wilma H. Trick,$^{2}$ Kareem El-Badry,$^{3}$ \newauthor Jan Rybizki$^{1}$ and Maosheng Xiang$^{1}$
\\
$^{1}$Max-Planck-Insitut f\"ur Astronomie, K\"oningstuhl 17, D-69117 Heidelberg, Germany\\
$^{2}$Max-Planck-Insitut f\"ur Astrophysik, Karl-Schwarzschild-Str. 1, D-85748 Garching
b. M\"unchen, Germany\\
$^{3}$Department of Astronomy and Theoretical Astrophysics Center, University of California Berkeley, Berkeley, CA 94720, USA
}
\date{Accepted XXX. Received YYY; in original form ZZZ}

\pubyear{2020}

\begin{document}
\label{firstpage}
\pagerange{\pageref{firstpage}--\pageref{lastpage}}
\maketitle

\begin{abstract}
Stars born at the same time in the same place should have formed from gas of the same element composition. But most stars subsequently disperse from their birth siblings, in orbit and orbital phase, becoming `field stars'. Here we explore and provide direct observational evidence for this process in the Milky Way disc, by quantifying the probability that orbit-similarity among stars implies indistinguishable metallicity. We define the orbit similarity among stars through their distance in action-angle space, $\Delta (J,\theta)$, and their abundance similarity simply by $\Delta$[Fe/H]. Analyzing a sample of main sequence stars from Gaia DR2 and LAMOST, we find an excess of pairs with the same metallicity ($\Delta\mathrm{[Fe/H]}<0.1$) that extends to remarkably large separations in $\Delta (J,\theta)$ that correspond to nearly 1~kpc distances. We assess the significance of this effect through a mock sample, drawn from a smooth and phase-mixed orbit distribution.  Through grouping such star pairs into associations with a friend-of-friends algorithm linked by $\Delta (J,\theta)$, we find 100s of mono-abundance groups with $\ge 3$ (to $\gtrsim 20$) members; these groups -- some clusters, some spread across the sky -- are over an order-of-magnitude more abundant than expected for a smooth phase-space distribution, suggesting that we are witnessing the \textquoteleft dissolution\textquoteright~of stellar birth associations into the field.
\end{abstract}



\section{Introduction}

The ever increasing amount of stellar spectra collected by spectroscopic Milky Way surveys, such as APOGEE \citep{apogee2}, GALAH \citep{sven}, LAMOST \citep{LAMOST1,LAMOST2}, RAVE \citep{2017ApJ...840...59C} amongst others, provides precise information on the element abundances of millions of stars. Combining these surveys with the second release of the Gaia satellite \citep{2018A&A...616A..11G}, that constrains the 6D phase space and orbit information for these stars, opens up the possibility to understand how our Galaxy has formed and evolved. 

 By studying the population properties of stars, their orbits, compositions, and ages we can learn about the assembly of different components of the Galaxy. Furthermore, clusters, either intact, dispersing or dissolving can teach us about the dynamical history of the Galaxy (e.g., \citealt{2012MNRAS.421.3338A,2013ApJ...764..124W,2015ApJ...807..104T}). 

One conceptual approach is the idea that stars that were born at the same time and in the same molecular cloud can reveal their common birth origin by their very similar chemical abundances \citep{2002ARA&A..40..487F}, even when they could have been dispersed into different places afterwards \citep{2015ApJ...807..104T, 2016ApJ...833..262H}. This is called \textquoteleft chemical tagging\textquoteright.

That stars disperse can mean that they are in different orbital phases (different locations) along nearly the same orbit. For example, when a star cluster gets disrupted, single-stellar population ``streams'' can be found extending for tens of kpc through the Galactic halo \citep{2014ApJ...795...95B,2017MNRAS.466.1741C}. Or this could mean that stars actually evolve to very different orbits, and this can happen through radial mixing or radial migration (e.g., \citealt{2002MNRAS.336..785S,2008ApJ...684L..79R,2015MNRAS.450.2354Q}). We now have clear and quantitave evidence that this migration is overall strong in the Galactic disc 
(e.g., \citealt{2018ApJ...865...96F}).

At late times (last 8 Gyrs), stars in the Milky Way were presumably born on disc-like orbits. However, discs are susceptible to perturbations, and fluctuations in the gravitational field cause a star to change its original orbital actions, or diffuse in action space (e.g., \citealt{2015ApJ...806..117F}). When radial migration is not at play, then we would expect that stars at a given radius have a clear relation between the age and metallicity of stars \citep{2002MNRAS.336..785S}, reflecting the successive enrichment of the birth gas. However observations have not shown such correlation in the solar neighborhood and instead have shown a large spread in metallicity, [Fe/H] \citep{1993A&A...275..101E, 2002A&A...394..927I}, implying that stars have in fact migrated over large radial distances during their lifetime \citep{2010ApJ...713..166B}. 

The task of identifying groups of stars from the same cluster purely by their chemical similarity, without information on velocity or distance, has been proposed \citep{2002ARA&A..40..487F} and put into practice by \citet{2016ApJ...833..262H,2017MNRAS.465..501S,2019A&A...629A..34G}, amongst others. In this scenario, for chemical tagging to be successful, one of the conditions is that the progenitor cloud is uniformly mixed before the first stars are formed \citep{2002ARA&A..40..487F}. In addition, birth clusters must have clear cluster-to-cluster abundance differences \citep{2016MNRAS.457.3934L}. Open clusters are good laboratories for testing whether these conditions hold \citep{2015A&A...577A..47B,2016MNRAS.463..696L, 2016ApJ...817...49B}. 

In recent years, data-driven methods have been used to extract high-precision abundances from spectra, even at moderate resolution and signal-to-noise \citep{2015ApJ...808...16N,2016ApJ...826L..25R,2017ApJ...843...32T,2019ApJ...879...69T}. However, pure chemical tagging is still a challenging technique \citep{2015ApJ...807..104T,2016sf2a.conf..333B}, as shown by the presence of \textit{doppelgangers} in field stars \citep{2018ApJ...853..198N}. 

Once precise abundances have been determined, a procedure is needed to identify potentially co-natal ``clumps'' in abundance space.
Some works make use of clustering algorithms such as {\sl k-means} (e.g, \citealt{2016ApJ...833..262H}) or the density-based spatial clustering of applications with noise (DBSCAN; \citealt{10.5555/3001460.3001507}) \citep[e.g.][]{2019ChA&A..43..225S,2019MNRAS.487..871P}. While for {\sl k-means} the number of clusters must be known in advance, and specified {\it a priori} in the algorithm, with DBSCAN the optimal number of clusters can be determined from the data in an automated way.  

Most stars formed in a molecular cloud are expected to disperse quickly, in $\lesssim$ 100 Myr \citep{2003ARA&A..41...57L}, in orbit and consequently in orbital phase on a longer timescale. However, their observable chemical abundances are expected to remain largely unchanged. Including  more dimensions than just chemical information (e.g. kinematics) increases the prospect of tracing back the origin of a dispersed cluster. Therefore exploring the extent to which stars with very similar abundances are also on similar or different orbits is a fundamental diagnostic. In the Galactic disc this tells us directly how strong radial migration was i.e., whether the present day orbit of normal disc stars has anything to do with their birth orbit. 

\citet{2019ApJ...884L..42K} has recently shown the existence of this dispersal by revealing that co-moving (in $\vec{x}, \vec{v}$) pairs in the solar neighborhood have a preference to have similar metallicities when compared to random field stars, even to distances beyond bound pairs. 
This opens up the possibility to find disrupting star clusters. However, a Cartesian coordinate system, as the one used in that work, may not be optimally suited to identify such signatures beyond the sun's vicinity.

Action-angles (J,$\theta$) are canonical coordinates to describe stellar orbits, and they may be a powerful coordinate system to find orbit-distribution sub-structure in our Galaxy. Whereas in configuration space ($\vec{x},\vec{v}$) each of the coordinates has a complex time evolution, in action-angle space the three actions are integrals of motion and constant, and the three angles evolve linearly with time \citep{binney2008}. Additionally, gradual changes in orbit may be described as a diffussion in action space \citep{2015MNRAS.449.3479S}. Hence, if we want to study larger volumes in the Milky Way (e.g, $d \gtrsim 200$\,pc, where the curvature of stellar orbits becomes pronounced), then this coordinate system may be better to identify stars that are on the same orbits, as compared to a Cartesian coordinate system ($X,Y,Z,U,V,W$). A cylindrical coordinate system ($R,\phi,z$) could also be used as a better spatial alternative over a larger region of the Galaxy.
Action-angles have already been used to study groups of stars on similar orbits, for example, \citet{2019MNRAS.484.3291T} have revealed rich orbital substructure in Gaia DR2, that extends over several kpc.  Action-angles are also convenient to study processes that might be responsible for orbit migration in the Galactic disc, like spiral arms \citep{2019MNRAS.484.3154S} and bars \citep{2019MNRAS.490.1026H,2019arXiv190604786T}.

Here we combine the spectroscopic information from LAMOST's latest data release, LMDR5 \citep{2019ApJS..245...34X} with the astrometric information from Gaia DR2 to investigate the probability that star pairs that are close in action-angle space have exceptionally similar metallicities, through $p(\Delta \mathrm{[Fe/H]}~|~\Delta (J,\theta))$. We start by defining a metric in action-angle space, combined with chemical information, in a \textit{generalized chemical tagging} approach. On this basis, we can show that the width of $p(\Delta \mathrm{[Fe/H]}~|~\Delta (J,\theta))$ grows continually with increasing $\Delta (J,\theta)$, from the regime of bound binaries to disc-halo pairs of stars, well beyond the distance regime probed in \citet{2019ApJ...884L..42K}. To see whether these ultra-wide pairs of stars trace the dispersal of birth associations, we then apply a friends-of-friends algorithm to stars of near-identical [Fe/H] to recover larger structures, recovering both known open clusters and widely dispersed groups. This method could constrain effects such as orbit diffusion in the Galactic disc. 

This paper is organised as follows: in Section 2 we present the data used in this study, observational and a mock catalog, in Section 3 the method: pairwise distances between stars, in Section 4 we present the results and analysis of the \textit{generalized chemical tagging}, in Section 5 the orbit clustering of stars with the same metallicities, Section 6 presents a comparison in ($\vec{r},\vec{v}$) configuration space; and finally Section 7 presents the summary followed by the appendix.

\section{Data}

\subsection{The Gaia DR2 $\otimes$ LAMOST DR5 Sample}
The analysis of this paper draws on the combination of the second Gaia data release, GDR2 \citep{2018A&A...616A..11G}, and the fifth data release (DR5) of the spectroscopic survey LAMOST (hereafter LMDR5) with stellar parameters derived from the Data-Driven Payne (DD-Payne, \mbox{\citet{2019ApJS..245...34X}}), which is a data driven model that includes constrains from theoretical spectral models to derive abundances. We obtain the positions (ra, dec), proper motions ($\mu_\text{ra}$, $\mu_\text{dec}$) and the parallaxes $\varpi$ from GDR2, where we impose selection criteria on the renormalized unit weight error $\leq$ 1.6 \citep{2018A&A...616A...2L} and on the parallax $\varpi > 0$. 

LAMOST provides spectra at a resolution of R $\sim$ 1800. We consider only stars with $\text{SNR}_{G} > 30$ in LMDR5 to decrease the uncertainties in $\text{[Fe/H]}$. For this subsample, the typical radial velocity precision is $(5-7)\,\rm km\,s^{-1}$, and the typical abundance precision is  $\sim 0.05-0.07$~dex for [Fe/H]. We make use of the spectroscopic parameters $T_\text{eff}, \log g, \text{[Fe/H]}$, and also the radial velocities. For this work we
make use of the recommended labels that combine results from the LAMOST-GALAH and LAMOST-APOGEE training sets, where we have selected stellar labels with no flags \citep{2019ApJS..245...34X}. 

Following \citet{2018MNRAS.481.2970C}, we calculate the spectro-photometric distances that combine the parallaxes and spectral information.  We consider main sequence (MS) stars with the following criteria: $4800 \text{K} < T_\text{eff} < 6000 \text{K}$, $\log g >$ 4.2. This selection of MS stars in $T_\text{eff}$ differs from the one adopted in \citep{2018MNRAS.481.2970C}, as estimates of [Fe/H] become less robust and accurate for $T_\text{eff}<4800$K. We then combine the dataset with \textit{2MASS} to obtain the \textit{K}-band magnitude, needed to apply our spectrophotometric distance model. Otherwise, we essentially follow here the model of \citet{2018MNRAS.481.2970C}, with further slight changes explained in more detail in Appendix~\ref{appendix_dist}. We are left with $\sim$ 550,000 MS stars after the GDR2$\otimes$LMDR5 cross-match and selection criteria. The stars in the sample here have distances up to 3 kpc, however the majority of them are at $d$ < 1.5 kpc.

\subsubsection{Wide Binaries in LAMOST as methodologial anchors}

In addition to our primary analysis of all possible pairs within the GDR2$\otimes$LMDR5 catalog, we also analyze a sample of 519 gravitationally bound wide binaries (WBs) for which both components have a high-quality spectrum from LAMOST. WBs represent the extreme low-$\Delta (J,\theta)$ limit for pairs close together in phase space: they not only have similar kinematics and in most cases formed from the same gas cloud, but they are still gravitationally bound. Because WBs are generally chemically homogeneous \citep[e.g.][]{Hawkins_2019}, the distribution of $\rm \Delta [Fe/H]$ within the WB sample represents the highest degree of chemical homogeneity we can expect to measure for stars formed at the same time and place within the Milky Way, given the noise properties of the GDR2$\otimes$LMDR5 sample. 

We select wide binaries using the same general procedure described in \citet{Elbadry_2018}: we identify pairs of stars with projected separations $s<50,000\,\rm AU$ that have parallaxes and proper motions consistent with bound Keplerian orbits and both have high-quality LAMOST spectra. The measured $\Delta(J,\theta)$ for WBs is necessarily low, but it is nonzero because (a) the nonzero orbital velocities cause the total space velocities of the components of WBs to differ at the $\sim 1\,\rm km\,s^{-1}$ level, and (b) uncertainties in the parallaxes and proper motions of both components inflate their $\Delta (J,\theta)$ to the noise floor. 

In contrast to the binary selection procedure of \citet{Elbadry_2018}, which relied only on 5D Gaia astrometry, we also make use of LAMOST radial velocities in our selection, requiring the radial velocities of the two components to be consistent within $2\sigma$ (Fig.~\ref{fig:WBs} in Appendix~\ref{sec:WBs_appendix}). This allows us to search for WBs out to a distance of 2\,kpc while maintaining a low contamination rate. We refer to \citet{ElBadry_2019} and \citet{Tian_2019} for detailed discussion of the wide binary selection procedure, contamination rate, and effective selection function. In this work, we restrict our analysis to the highest-quality subsample of the WBs: those which both components have a LAMOST spectrum with ${\rm SNR}_G > 50$ and precise Gaia astrometry ($\varpi/\sigma_{\varpi}>10$). 
\subsection{A Mock Catalog with a Smooth and Phase-Mixed Orbit Distribution}
\label{sec:gaia_mock}

As a null hypothesis for our analysis, we need to understand the amount of clustering we expect to find in the case that all stars are in a smooth orbit distribution fully phase-mixed, where [Fe/H] only changes gradually with the ``Galactic component'', or radius.

We do this by creating a mock observation that matches our GDR2$\otimes$LMDR5 selection in volume and depth, based on the Gaia DR2 mock stellar catalog by \citet{2018PASP..130g4101R}. This catalog was created using a chemo-dynamical model based on Galaxia \citep{2011ApJ...730....3S}, where the stars are sampled from the Besan\c{c}on Galactic model \citep{2003A&A...409..523R}. 
The 2003 Besan\c{c}on model prescribes smooth distributions in phase and abundance-space to the four main Galactic components (thin, thick-disc,bulge and halo), with basic observational constraints, like the age-velocity-relation, age metallicity distribution and radial metallicity gradient, imprinted. It should be noted that the sampled version in GDR2 mock neither includes binaries nor spiral arms (any
localised/clumpy star formation).
We select stars in this GDR2 mock with criteria resembling those of our dataset: in $T_\text{eff}$ and $\log g$, with additional cuts in parallax and magnitude: $\sigma_{\varpi}/\varpi < 0.1$ and 10$<$  \texttt{phot\_g\_mean\_mag} $<$ 14. The values provided in the catalog are noise-free, hence parallaxes could be directly inverted to give exact model distances \citep{2018PASP..130g4101R}. We proceed to add noise to the parallax by sampling from a Gaussian with the true value of $\varpi$ as mean and $\sigma_{\varpi}$ as the standard deviation, as suggested in \citet{2018PASP..130g4101R}. 

\begin{figure}
\centering
\includegraphics[width=\columnwidth,trim={0 3.5cm 0 3cm},clip]{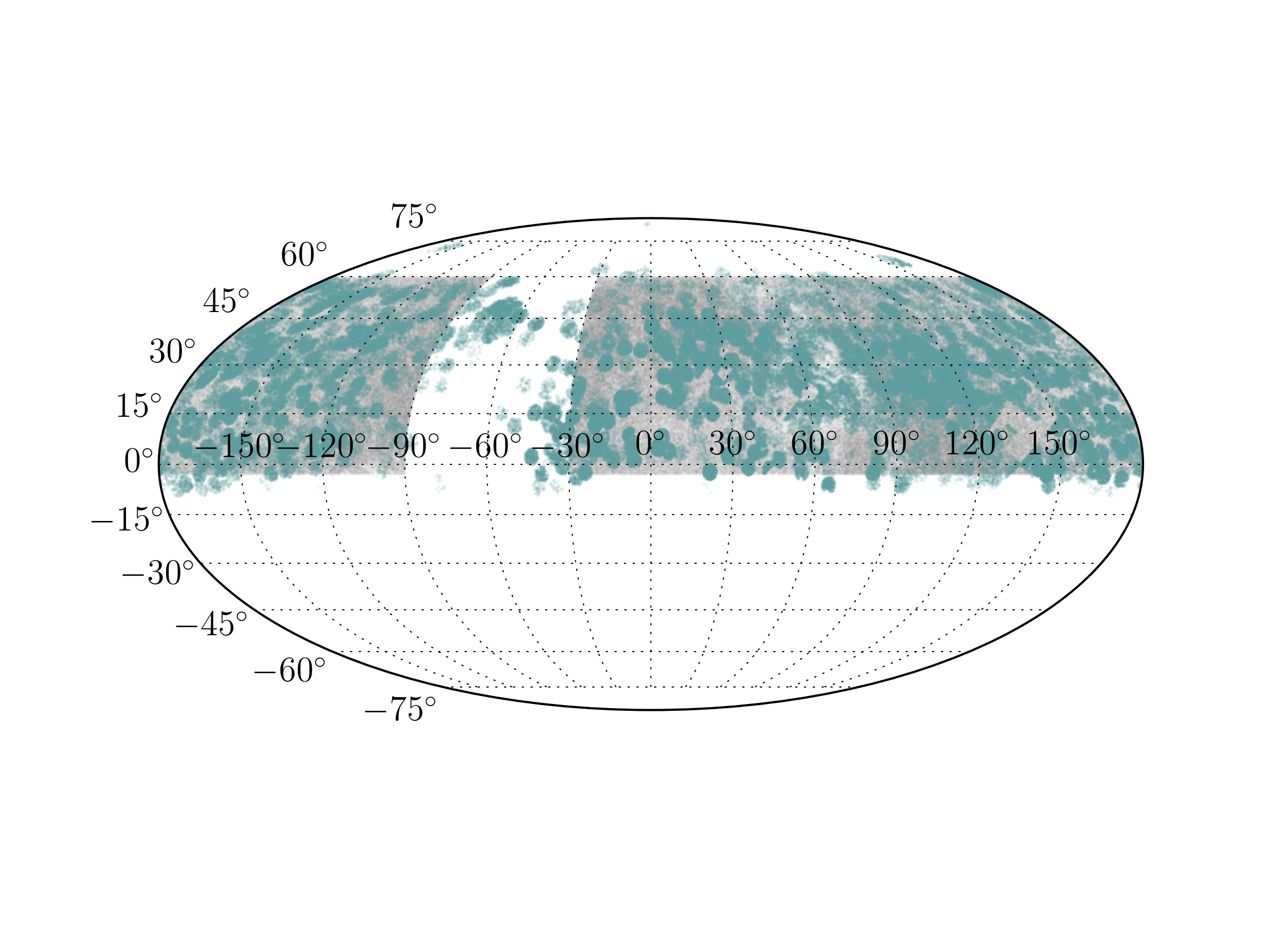}
\caption{Footprint on the sky in equatorial coordinates of the LMDR5 sample in blue dots, and the mock catalog in grey. Here we can see that the LAMOST survey covers the northern hemisphere, and we select the stars in the mock catalog accordingly, considering the areas of the sky that were mostly completely covered with LMDR5 stars. }
\label{fig:foot_coverage}
\end{figure}

Then we match the sky coverage of the LAMOST survey, which covers much of the northern sky (Fig.~\ref{fig:foot_coverage}). After applying all of these cuts, we are left with $\sim$ 580,000 stars in the GDR2 mock, matching the sample size of our GDR2$\otimes$LMDR5 dataset. 

\section{Methodology:  Pairwise Distances in Actions, Angles, and [Fe/H]}
\label{sec:pairwise_dist}
In order to see if we can find signatures of stars that were born at the same time from the same material, we investigate if we can quantify statistically how much closer pairs of stars are in [Fe/H] if they are close in orbit space by studying $p(\Delta \mathrm{[Fe/H]}~|~\Delta (J,\theta))$. 

\subsection{Choice of variables: $\Delta (J,\theta)$ and $\Delta$[Fe/H] }

In the following sub-sections we describe operationally how to define and then calculate distances between pairs of stars, both in orbit space and in abundance space. 

For two stars that are on nearly the same orbit and nearly the same orbital phase, their distance 
can be well defined in the Cartesian configuration space $(\vec{x},\vec{v})$, as done by \citet{2019ApJ...884L..42K}. For wider separations, one could use  classical integrals $E$ and $L_{z}$ \citep{1916MNRAS..76..552J,1963AJ.....68....1C,1965ARA&A...3..113O, 1984MNRAS.206..159B}.
But action-angle variables ($J,\theta$) are arguably the best set of coordinates, as they form a 6D canonical coordinate system with several advantages. For an axisymmetric gravitational potential, all three actions $(J_{R}, J_{z}, J_{\phi}$) are integrals of motion \citep{2012MNRAS.426.1324B}, where $J_{R}$ quantifies the oscillations of the orbit inwards and outwards in the radial direction, $J_{z}$ quantifies the oscillations in the vertical direction and $J_{\phi}$ (or $L_{z}$) is the azimuthal action that equals the angular momentum in the $z$ direction. All actions have the same units, kpc$\times$km/s.
Actions are complemented by their three corresponding angles: $\theta_{R}$, $\theta_{z}$ and $\theta_{\phi}$. These angles, reflecting the orbital phase in these coordinates, increase linearly with time, in practice modulo $2\pi$. If a system is fully phased mixed, then the angles should be uniformly distributed between 0 and 2$\pi$.

In the presence of non-axisymmetric structures such as spiral arms or a bar, the three actions $J_{R}, J_{z}, J_{\phi}$ are not well defined, and are not exactly integrals of motion. However, (axisymmetric) approximations can still be made to compute them. 
For a thorough description of action angle variables, we redirect the reader to Section 3.5 of \citet{binney2008}.

\subsection{Action-angle computation}

The calculation of actions and angles requires both phase-space coordinates, and an (assumed) gravitational potential. 
If we assume that the Galaxy's potential is close to an axisymmetric St{\"a}ckel potential, then the actions and angles can be easily calculated. We make use of the python package \texttt{galpy}, with its implementation of the action estimation algorithm \emph{St{\"a}ckel fudge} \citep{2012MNRAS.426.1324B} along with the \texttt{MWPotential2014} model. The latter considers a simple axisymmetric Milky Way potential model with a circular velocity of 220 km/s at the solar radius of 8 kpc \citep{galpy}. Note that the absolute values of the actions never enter the subsequent analysis, just their differences. So, the choice of an updated circular velocity (e.g., \citet{2019ApJ...871..120E}) would not significantly alter the results.

 For the location and velocity of the Sun within the Galaxy we assume (X,Y,Z) = (8,0,0.025) kpc and (U,V,W)$_{\odot}$ = (11.1,12.24,7.25) km/s \citep{2010MNRAS.403.1829S} to first calculate Galactocentric coordinates and then actions from the observed (ra, dec, $d$, $v_{\rm los}$, $\mu_\text{ra}$,$\mu_\text{dec}$) of each star. As noted by \citet{2018MNRAS.481.2970C}, the largest contribution to the action uncertainties comes from the distances. However, by calculating the spectrophotometric distances as described in Appendix ~\ref{appendix_dist}, we obtain improved distances (at least for distant stars) with uncertainties of $\sim$ 7\% for single stars. We refer the reader to Section 5.2.1 in \citet{2018MNRAS.481.2970C} to see the extent of the uncertainties in action space when applying this spectrophotometric distance model. Typical action uncertainties are $\sim 5-8$\%. 

\subsection{Defining a Metric in Action space}

To calculate the pairwise distances between stars in action space, we first must define a metric that combines the three actions $J_{R}, J_{z}$ and $J_{\phi}$. For subsequent combination with the angle separation metric, we want this metric---or distance---to be unitless. Therefore we normalise each dimension by the ensemble variance in each quantity, defining the distance between a pair of stars (i,j) in action space as

\begin{equation}
\Delta J^{2}_{ij} \equiv w_{J_R}\cdot (J_{R,i} - J_{R,j})^{2} + w_{J_z}\cdot
(J_{z,i} - J_{z,j})^{2} + \\ w_{J_\phi}\cdot 
(J_{\phi,i} - J_{\phi,j})^{2}
\label{eqn:1}
\end{equation}

with
\begin{equation}
\begin{split}
w_{J_k} \equiv \frac{1}{\mbox{Var}(J_{k})}, k \in \{R,\phi,z \}
\end{split}
\label{eqn:2}
\end{equation}
where the variance is defined as Var = $\sum_{i=1}^N (x_{i}-\bar{x})^{2}/(N-1)$ for a sample size $N$.

\subsubsection{Defining a Metric in Action-Angle space}

Stars that drift apart in orbit space will then also drift apart in orbital phase, unless the orbital frequency stays identical. Therefore, the distance between two stars in orbital phase, or angle, matters. We define an analogous 6D metric that combines the actions $J_{R}$, $J_{z}$, $J_{\phi}$ with their respective angles $\theta_{R}$, $\theta_{z}$ and $\theta_{\phi}$.

We start with the angle part: 
\begin{equation}
\begin{split}
\Delta \theta^{2}_{ij}  \equiv w_{\theta_R}\cdot \Delta \theta^{2}_{R,ij} + w_{\theta_z}\cdot \Delta \theta^{2}_{z,ij} + w_{\theta_\phi}\cdot \Delta \theta^{2}_{\phi,ij} .
\end{split}
\label{eqn:3}
\end{equation}
$w_{\theta_k} \equiv \frac{1}{\mbox{Var}(\theta_{k})}, k \in \{R,\phi,z \}$, with 
\begin{equation}
\Delta \theta_{k,ij} \equiv \mathrm{min} \left[|\theta_{k,i} - \theta_{k,j}|, ( 2\pi - |\theta_{k,i} - \theta_{k,j}|)\right], k \in \{R,\phi,z \} \label{eq:periodic_angle_distance}
\end{equation}

By definition, the angles $\theta_k$ are in the range and periodic in $[0,2\pi]$. Eq.~\ref{eq:periodic_angle_distance} ensures that the correct (and smallest) angle distance is used. Again, we introduce a normalisation factor $w_{\theta}$ for each of the angles. We note that the variance in $\theta_{R}$ and $\theta_{z}$ has roughly the same value, we can see stars in basically all phases of their vertical and radial oscillation. For the azimuthal direction, defined to be 0 at the line from the Sun to the Galactic center, only a small fraction of angles will be within the sample volume. The values that we considered for the variance are the typical distance two stars can have in the angles. We use the same weights for the real data and the mock catalog. 
Combined, this yields a sensible action-angle distance metric:

\begin{equation}
\Delta(\mathrm{J, \theta})^{2}_{ij} \equiv \Delta J^{2}_{ij} + \Delta \theta^{2}_{ij},
\label{eqn:4}
\end{equation}
where both components of the metric are unitless. 

\subsection{Distance in Abundance Space: $\Delta$[Fe/H]}

We define a distance in abundance space by considering the differences in [Fe/H] exclusively. This is for several reasons: [Fe/H] has the largest variance compared to [X/Fe], it is robustly determined and it is available also in the mock catalog. Then, the pairwise distance for the metallicity is defined as:

\begin{equation} 
\Delta_{ij} \mbox{[Fe/H]} \equiv |\mbox{[Fe/H]}_{i} - \mbox{[Fe/H]}_{j}| 
\label{eqn:5}
\end{equation}
For our dataset, the uncertainties in [Fe/H] are less than 0.1 dex. 

\begin{figure}

\includegraphics[width=\columnwidth]{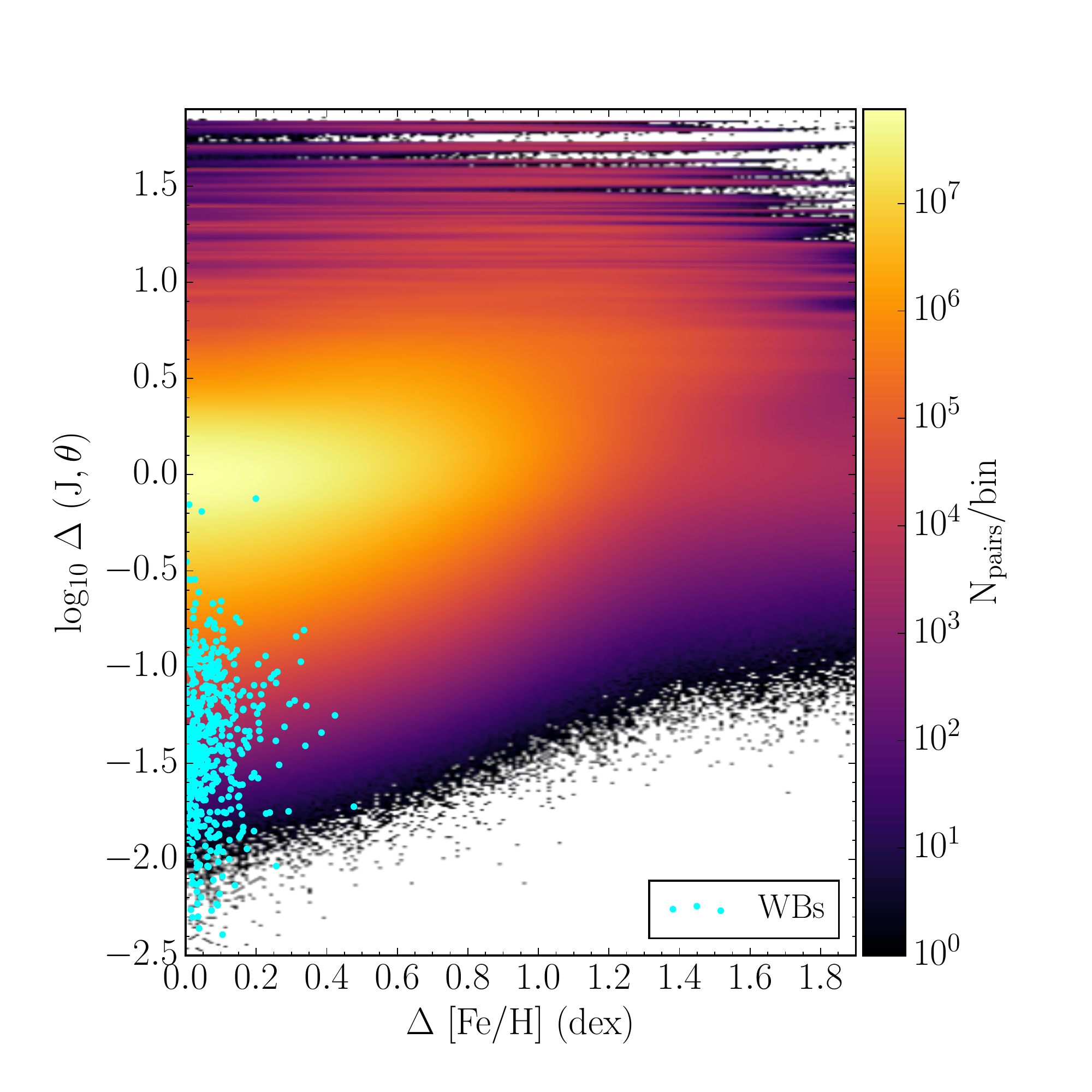}
\caption{Distribution of pairwise distances in action-angle and metallicity space for MS stars in LAMOST $\times$ Gaia DR2, as defined in Eqs.~\ref{eqn:3},~\ref{eq:periodic_angle_distance}, ~\ref{eqn:4} and ~\ref{eqn:5}. In cyan dots we show how Wide Binary pairs (WBs) are distributed in this same space. WBs are mostly concentrated at small distances in both action-angle, log$_{10} \Delta (J,\theta)$ space and metallicity $\Delta$[Fe/H], as expected from stars that were born together. There is a smooth transition from stars close in [Fe/H] - ($J, \theta$) towards stars at larger $\mbox{[Fe/H]}$ and $(J, \theta$) distance. The typical uncertainties in $\Delta(J,\theta)$ are $\sim$ 5\%. The bin size is 0.01 in this figure.}  
\label{fig:hist6d}
\end{figure}

With these definitions, we can proceed to explore the action-angle and metallicity distances between pairs of stars, as illustrated in Fig.~\ref{fig:hist6d}, where we show the complete distribution of the pairwise distances $p(\Delta \mathrm{[Fe/H]}~|~\Delta (J,\theta))$ that we obtain. 

\section{Generalized chemical tagging analysis: Orbit-similarity vs. Abundance-Similarity} 

\label{sec:results}
In Sec.~\ref{sec:pairwise_dist} we have defined pairwise distances between stars, both in action space only and in the full action-angle space. We will mainly analyse the results from a 6D phase-space metric, combined with chemical information. However, in Appendix~\ref{sec:appendix_action}, we also show the results for the distance in action only space (i.e., a 3D coordinate system) also combined with [Fe/H].

\subsection{Abundance differences of stars on similar orbits: $p\bigl (\Delta\mathrm{[Fe/H]}~|~\log_{10}\Delta (J,\theta)\bigr  )$ }

In Fig.~\ref{fig:hist6d}, we present the distribution of distances in action-angle space (using the metric defined in Eqs.~\ref{eqn:3},~\ref{eq:periodic_angle_distance},~\ref{eqn:4}) vs. $\Delta$[Fe/H] for all $\sim 10^{11}$ stellar pairs in our sample. The peak of the distribution is reached at $\sim$ log$_{10}\Delta (J,\theta) = 0$ which by construction is the mean pair separation. We have additionally divided the distances by the number of dimensions, 6 in action-angle space and by 3 in the action only case. This figure already illustrates the broad trend that stars close in $\Delta (J,\theta)$ tend to be close in $\Delta$[Fe/H] and {\it vice versa}. The extremes are wide binaries (bottom left of Fig.~\ref{fig:hist6d}) and presumably disc-halo pairs (top right of Fig.~\ref{fig:hist6d}). Those latter pairs would differ in both chemical composition and they would also be in completely different orbits. Overall, this shows that stars that are on similar orbits and close in the phase angles have also similar metallicities. In this figure we also show the distribution of WBs in action-angle and metallicity space. We discuss in more detail this sample in Appendix~\ref{sec:WBs_appendix}.

To quantify this effect and put it into perspective, we compare the distribution of $p\bigl (\Delta\mathrm{[Fe/H]}~|~\log_{10}\Delta (J,\theta)\bigr  )$ for the observed in data to an idealized mock galaxy, that has broad population gradients, but no clustered star-formation. As mentioned in Sec.~\ref{sec:gaia_mock}, we make use of the Gaia DR2 mock stellar catalog by \citet{2018PASP..130g4101R}. 

\begin{figure*}
\centering
\hspace*{-0.2cm}\includegraphics[width=.45\textwidth]{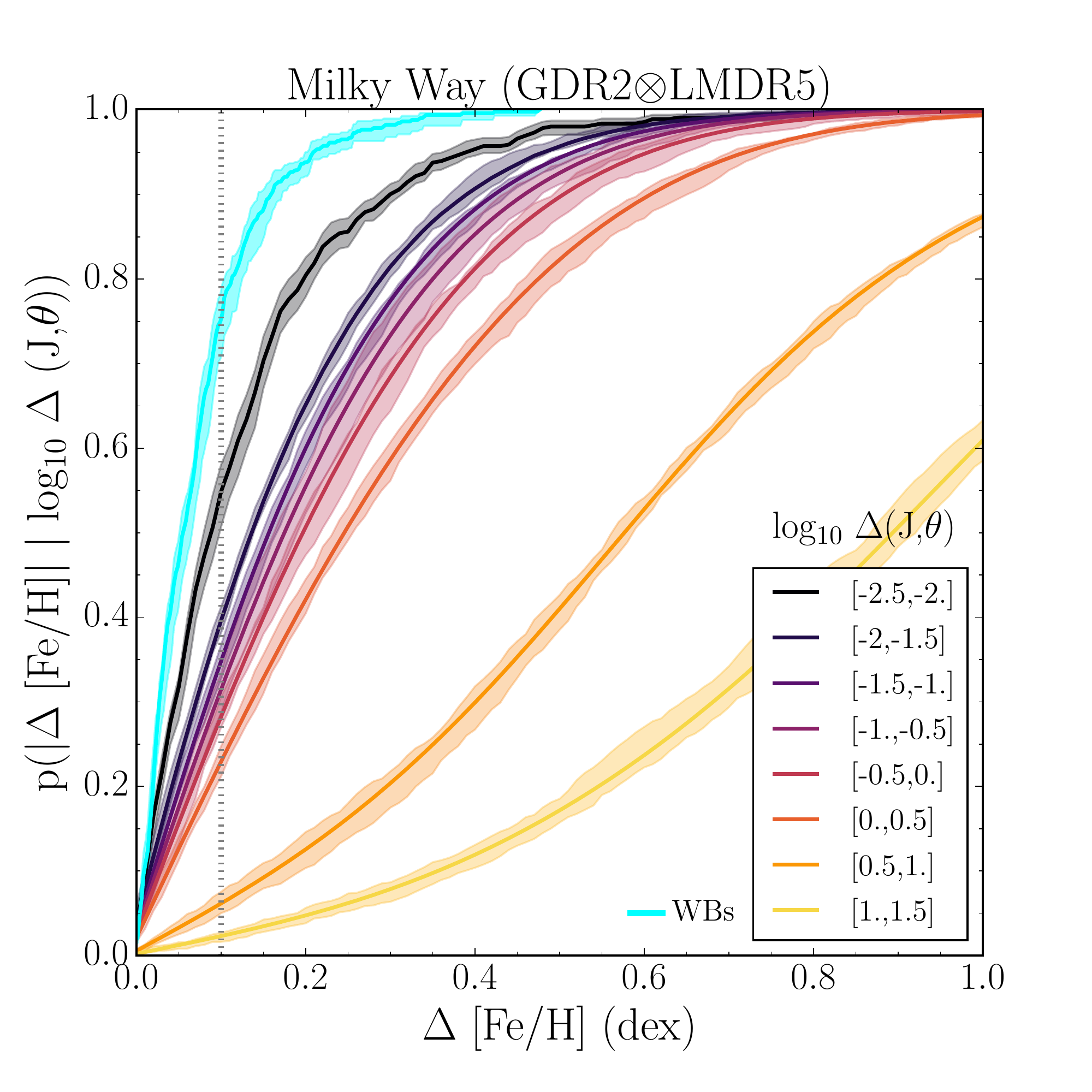}\hspace*{-0.2cm}
\hspace*{-0.2cm}\includegraphics[width=.45\textwidth]{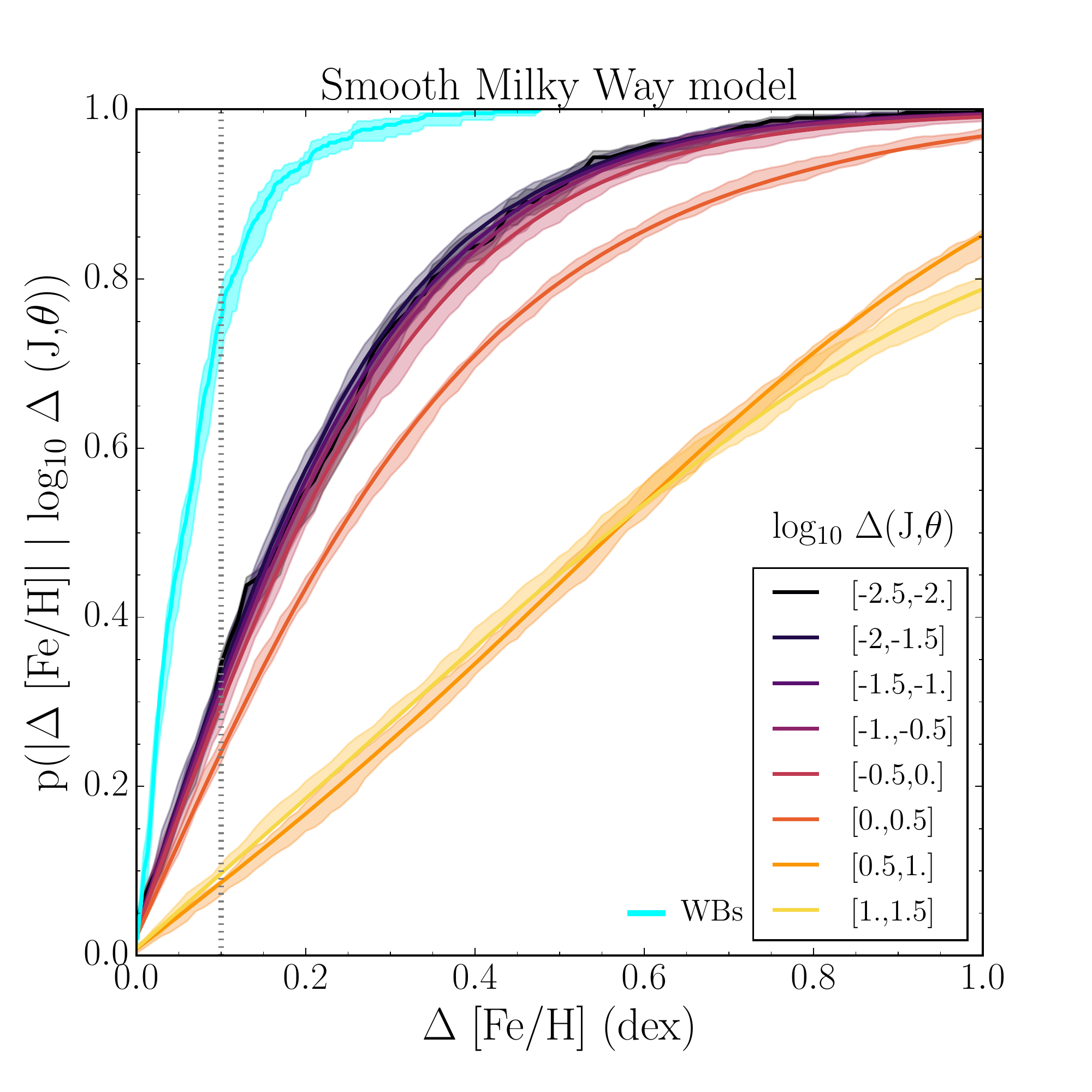}\hspace*{-0.2cm}\\

\caption{Correlation between $\Delta (J, \theta)$ and $\Delta$[Fe/H] in the Milky Way vs. a mock Galaxy with no clustered star formation. Here we show the CDF of pairs in given distance bins $\Delta (J, \theta)$ as a function of $\Delta$ [Fe/H], for the LMDR5 $\times$ Gaia DR2 MS stars in the left, and the mock catalog in the right. The width of these lines show the 5th and the 95th percentile of a bootstrap re-sampling. The cyan line shows the WBs, for comparison. The dashed line is located at $\Delta$[Fe/H] = 0.1, that we consider as an upper limit for the uncertainties in [Fe/H]. We observe that for the first bin---with the smallest log$_{10} \Delta (J, \theta$) (black line)--- $\sim $60\% of pairs with action-angle distances log$_{10} \Delta (J, \theta$) < -2 have metallicity within the uncertainty. As for the mock catalog, we see that for the first 5 bins, the lines are located at almost the same position, and we find a smaller value for p($\Delta$[Fe/H] $|$ log$_{10} \Delta (J, \theta$)) for the smallest log$_{10} \Delta (J, \theta$) than the one shown by the data.}

\label{fig:dist_action_angles_mock}
\end{figure*}

In Fig.~\ref{fig:dist_action_angles_mock}, we present the cumulative distribution function (CDF) of stars as a function of $\Delta$[Fe/H]. The left side of this figure shows the GDR2$\otimes$LMDR5 dataset (that we will now call the real MW pairs) where each coloured line represents the CDF for different orbit-similarity bins in Fig.~\ref{fig:hist6d} separated by 0.5 in log$_{10}\Delta (J,\theta)$. The right panel in Fig.~\ref{fig:dist_action_angles_mock} shows the same but for the mock data pairs: Following the same procedure as for the real MW pairs, we calculated the metric in action-angle and [Fe/H] space for the mock data pairs, with Eqs.~\ref{eqn:3}, ~\ref{eq:periodic_angle_distance}, ~\ref{eqn:4} and ~\ref{eqn:5}. We use the same values for the variance (Eq.~\ref{eqn:2}) that we obtain from the real MW pairs, for the mock ones given that their values are similar. We obtain a histogram in log$_{10}\Delta (J,\theta)$-$\Delta$[Fe/H], and then using the same bins as for the real MW pairs, we produce the CDF. In both figures the cyan line shows the complete CDF of the WB pairs as presented by cyan dots in Fig.~\ref{fig:hist6d}, where the WBs have $p(\Delta$  [Fe/H] = 0.1) $\sim$ 0.8. This clearly shows that most of the distribution of WBs is in fact close in [Fe/H].

For the real MW pairs, we find that in the smallest bin in log$_{10}\Delta (J,\theta)$, $\sim$ 60\% of the pairs have metallicity differences within the measurement uncertainty of 0.1 dex. As for the mock data pairs, we find that the smallest bin in action-angle distance has $\sim$ 40\% of pairs at 0.1 dex in $\Delta$[Fe/H]. Given the large sample sizes, these differences are highly significant. Most importantly, the CDF's in the five closest orbit bins in the mock data pairs are nearly identical, but the fraction of pairs with indistinguishable $\Delta$[Fe/H] rises towards small $\Delta (J,\theta)$ in the real MW pairs. As for the large $\Delta (J, \theta$) bins in the real MW pairs, the separation between them becomes wider, this is because in this regime we would expect to find more random pairs, that are not actually physically related. They are not only far apart in the $\Delta (J, \theta$) metric, but also in [Fe/H]. The highest $\Delta (J, \theta$) bin in the mock data pairs does not show a strong difference in [Fe/H] as the one observed in the real MW pairs, this is because of how the different components (thick disc, halo and bulge) are simulated in that catalog \citep{2018PASP..130g4101R}.

\begin{figure*}
\centering
\hspace*{-0.2cm}\includegraphics[width=.45\textwidth]{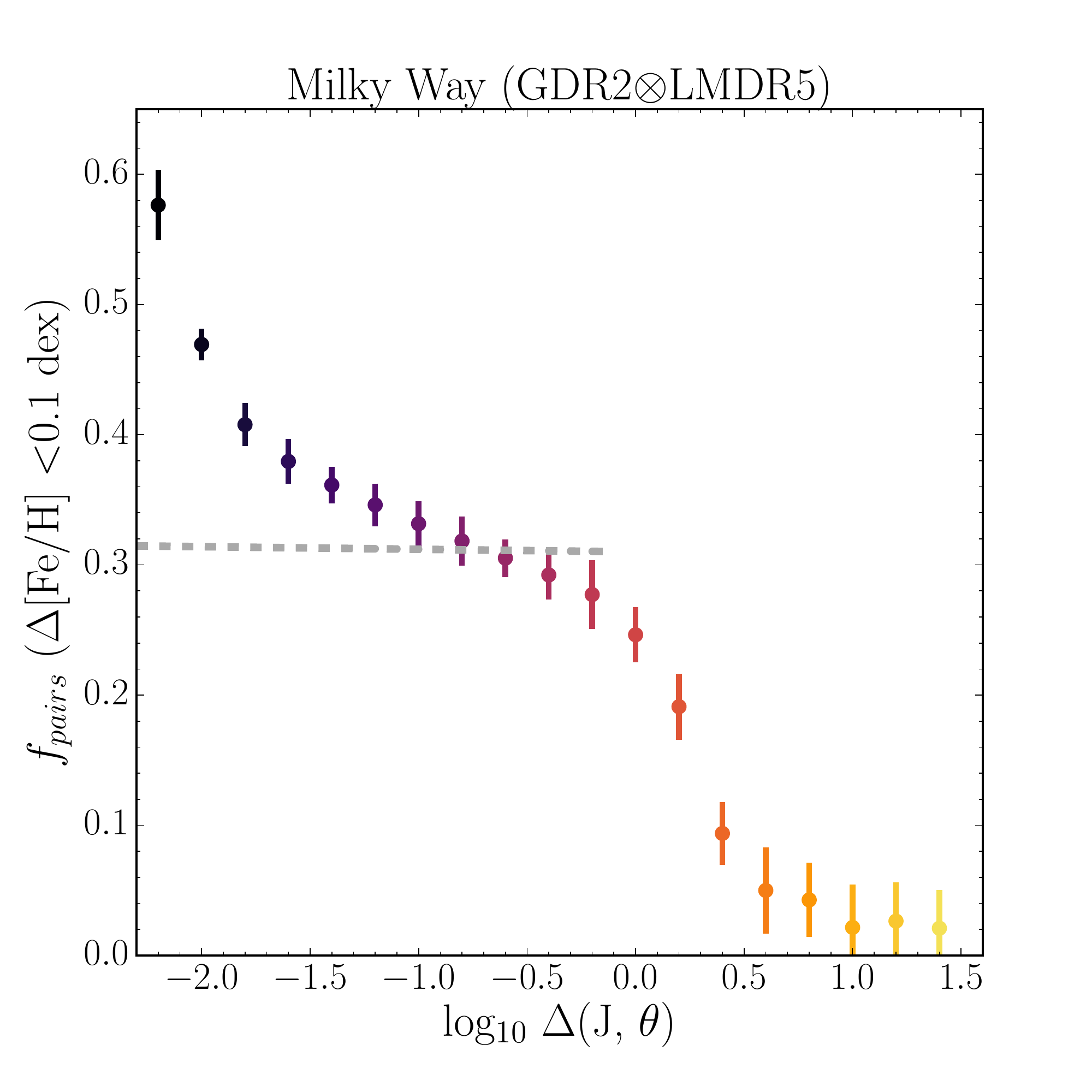}\hspace*{-0.2cm}
\hspace*{-0.2cm}\includegraphics[width=.45\textwidth]{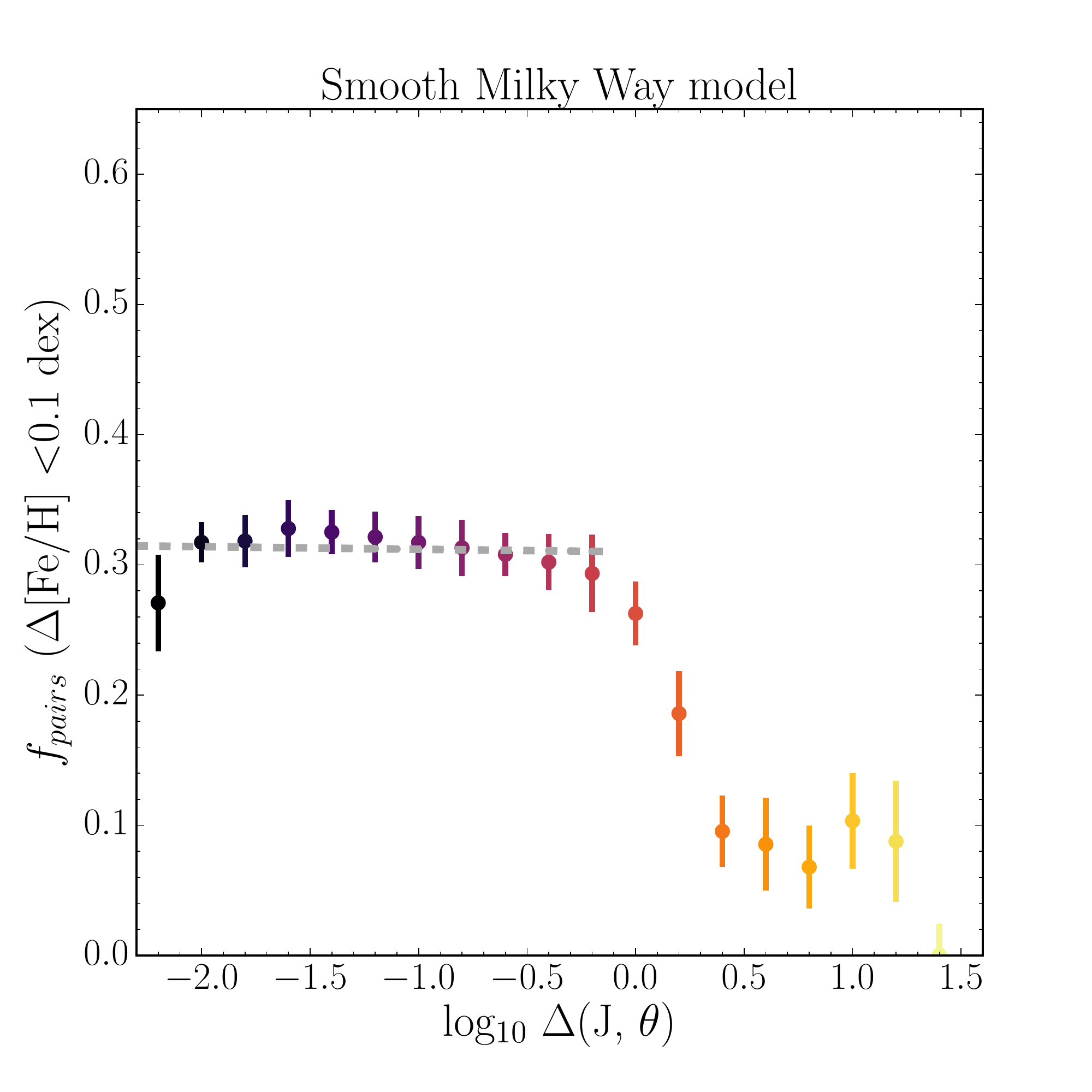}\hspace*{-0.2cm}
\caption{Fraction of pairs with indistinguishable metallicities ($\Delta$[Fe/H]$<0.1$ dex) at different bins in log$_{10} \Delta (J, \theta$) for the MW pairs to the left, and the mock pairs to the right. Each colored dot corresponds to bins of 0.2 in log$_{10} \Delta (J,\theta$). The colors in this plot indicate different bin values, similarly to Fig.~\ref{fig:dist_action_angles_mock}.
This clearly illustrates the differences between mock and MW pairs, where we find that the fraction of mock pairs is roughly flat ($\sim$ 31\%) for these 9 bins. Whereas for the MW pairs we see that for the first 2 bins the fraction of pairs is $\sim$ 60\% and $\sim$ 45\% respectively. The grey dashed line is a fit to 9 bins in the mock pairs --excluding the first bin-- that we then over-plot also in the left panel with the MW pairs. Therefore, we find that there is a large fraction of real MW pairs at small $\Delta (J, \theta$) with similar metallicities when compared to the mock pairs. In both mock and MW pairs the fraction of pairs decreases below 1\% for the last two bins in log$_{10} \Delta (J, \theta$). } 
\label{fig:dist_action_angles_mock2}
\end{figure*}

\subsection{The fraction of stars with the same [Fe/H], as a function of orbit similarity: $f_{pairs}\bigl (\log_{10}\Delta (J,\theta)~|~\Delta\mathrm{[Fe/H]}<0.1\bigr  )$ }

\begin{figure*}
\centering

\hspace*{-0.2cm}\includegraphics[width=.45\textwidth]{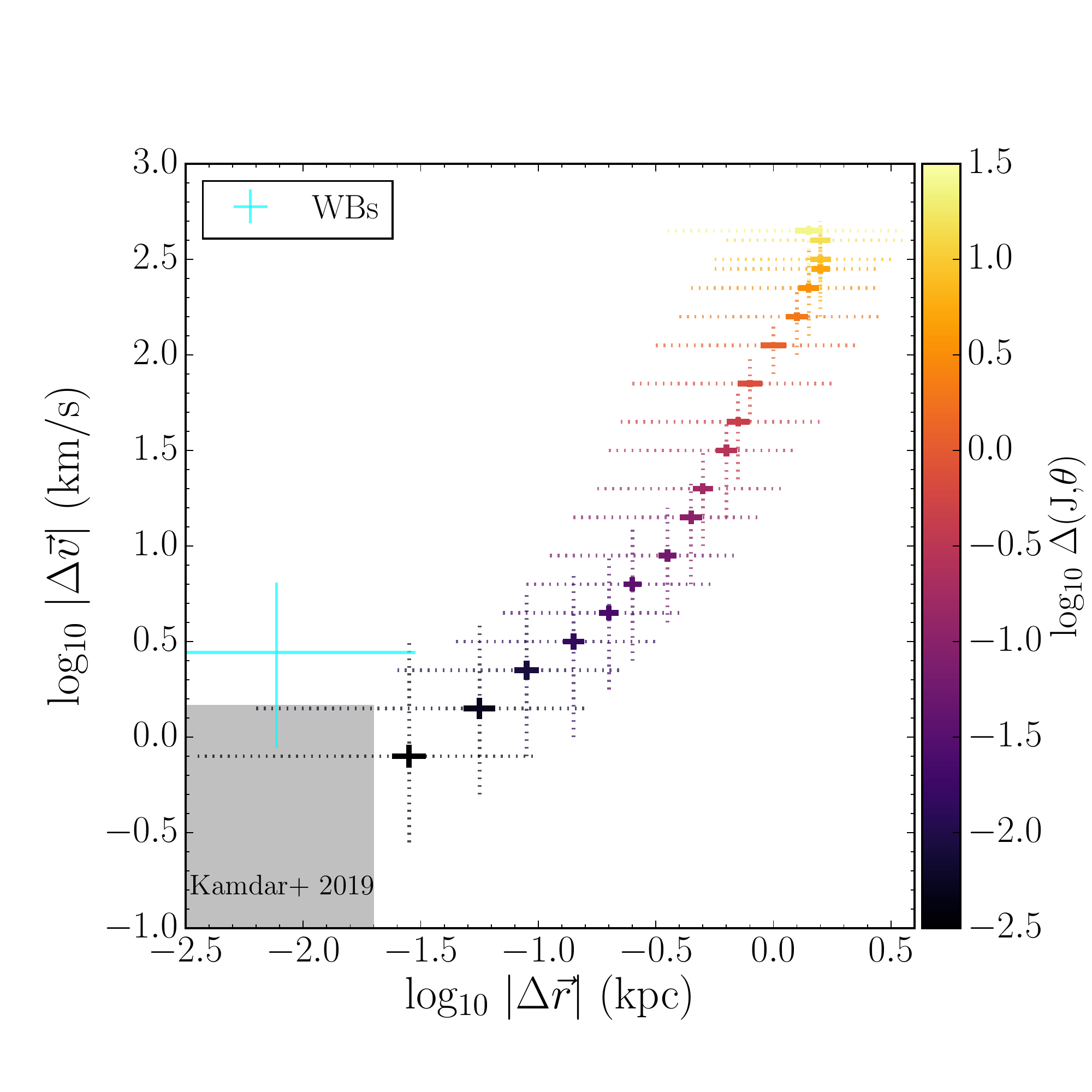}\hspace*{-0.2cm}
\hspace*{-0.2cm}\includegraphics[width=.45\textwidth]{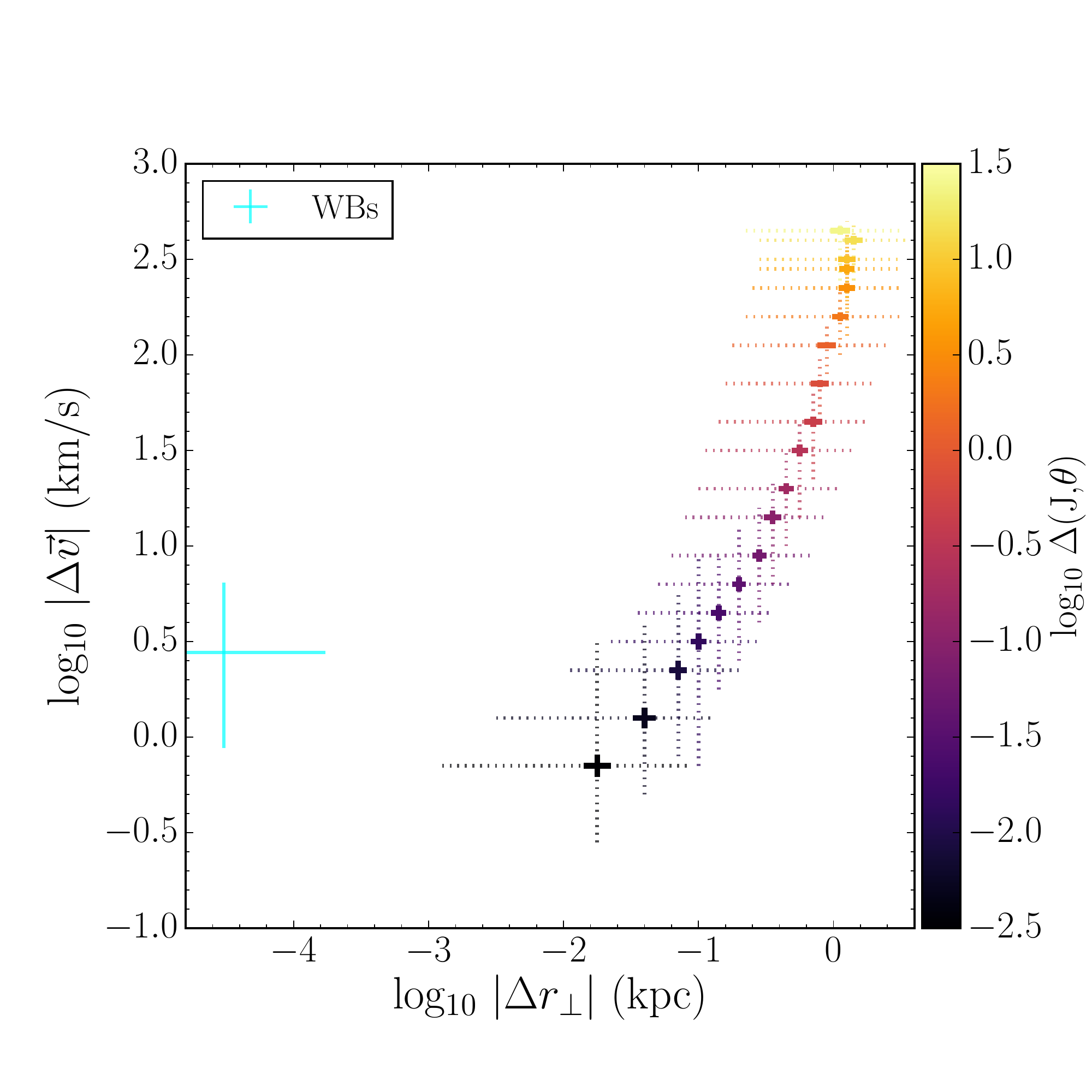}\hspace*{-0.2cm}\\
\caption{Differences in 3D velocities $\Delta\vec{v}$ and positions $\Delta\vec{r}$ to the left and transverse distance $\Delta {r_{\perp}}$ to the right; both in bins of 0.2 in log$_{10} \Delta (J, \theta$) for our MW pairs. In both panels we show the WBs in cyan, that are located at small $\Delta\vec{v}$, $\Delta\vec{r}$ and $\Delta {r_{\perp}}$, respectively. For the bins, the solid lines show the uncertainty of the mean value (calculated via bootstrapping), the dashed line --and solid line in the WBs-- show the 5th and 95th percentile. To the left, the smallest difference in log$_{10} \Delta (J, \theta$) is located also at small differences in velocity-distance space, as expected, given that ($J, \theta$) is only a different coordinate system for the same phase-space as ($\vec{r},\vec{v}$). To the right, the WBs have a projected --or transverse-- distance of $\sim$ 0.03 pc whereas the smallest bin in $\Delta (J, \theta$) is well beyond the WB regime, showing a mean value of $\sim$ 10 pc. 
Finally, the left side of this figure illustrates that using distances in action-angle space is analogous to position-velocities. The grey rectangle shows \citet{2019ApJ...884L..42K} co-moving pair selection in $\Delta\vec{v}$ and $\Delta\vec{r}$.} 

\label{fig:vels_x_mock}
\end{figure*}

We now consider a statistic that perhaps speaks more immediately to the question of whether we see birth associations of stars disperse and transition of field stars. Specifically we consider the fraction of pairs at a given $\log_{10}\Delta (J,\theta)$ that have indistinguishable [Fe/H], $f_{pairs}\bigl (\log_{10}\Delta (J,\theta)~|~\Delta\mathrm{[Fe/H]}<0.1\bigr)$. As we consider larger orbit separations $\log_{10}\Delta (J,\theta)$
the chances of finding pairs of different birth origin should increase, and $f_{pair}(\Delta\mathrm{[Fe/H]}<0.1)$ should decrease. We choose $\Delta\mathrm{[Fe/H]}<0.1$ to denote indistinguishable [Fe/H] as our individual metallicity precision is about 0.07~dex.
But of course, given the (local) metallicity dispersion of the low-$\alpha$ disk, the condition $\Delta\mathrm{[Fe/H]}<0.1$ may be satisfied for many star pairs born at different times in different parts of the disc. Such a test can therefore be only `statistical', and we again put our findings into perspective by comparison with a mock catalog from a smooth galaxy model (with population gradients).
The result of this analysis is quite striking, and is summarized in Fig.~\ref{fig:dist_action_angles_mock2}. The panels show $f_{pairs}(\Delta\mathrm{[Fe/H]}<0.1 )$ as a function of $\log_{10}\Delta (J,\theta)$ (in bins of 0.2); the left panel shows the observations, the right panel the smooth mock catalog. The majority of real MW pairs in the closest
$\log_{10}\Delta (J,\theta)$-bin have indistinguishable [Fe/H], which then decline to $\sim$30\% at
$\log_{10}\Delta (J,\theta)\sim 0$, and then quite precipitously fall to nearly 0 at $\log_{10}\Delta (J,\theta)> 0.8$ (presumed disc-halo pairs).
The right panel, with the analogous analysis from the smooth galaxy model, shows a qualitatively similar behaviour at $\log_{10}\Delta (J,\theta)> 0$. But there is a striking difference for $\log_{10}\Delta (J,\theta)< -0.5$: the fraction of mono-abundance pairs is constant for all smaller $\log_{10}\Delta (J,\theta)$, while the fraction rises steeply for the actual observations, where we see that the fraction of MW pairs for the first two bins is $\sim$ 60\% and 45\% respectively. Even though the mock galaxy is based on a chemo-dynamical model, we would not expect to find pairs of stars clumped in action-angle and [Fe/H]. Because the stars in GDR2 mock are distributed smoothly in phase-space there is no clustering \citep{2018PASP..130g4101R}. This means that there is a distinct excess of mono-abundance pairs at small orbit-separations in the real data, just as expected if there is a decreasing fraction of birth pairs as $\log_{10}\Delta (J,\theta)$ increases. On the scales of parsecs, this effect has been seen before \citep{Oh2017,2019ApJ...884L..42K}. But we now see this effect in our data to far larger distances. This may not be apparent from the X-axis $\log_{10}\Delta (J,\theta)$; therefore we illustrate in Fig.~\ref{fig:vels_x_mock} how a certain $\log_{10}\Delta (J,\theta)$ translates into typical spatial distances [in pc] or velocities [km/s]. Fig.~\ref{fig:vels_x_mock} takes the same bins in $\log_{10}\Delta (J,\theta)$ (in the same color-coding as in Fig.~\ref{fig:dist_action_angles_mock2}) and calculates for these pairs the mean $\Delta\vec{r}$ and $\Delta\vec{v}$. The differences between the real and smooth mock data in Fig.~\ref{fig:dist_action_angles_mock2} in the first 6 or 7 bins, in Fig.~\ref{fig:vels_x_mock} now informs us that this corresponds to 10~km/s and nearly 500~pc. In the right panel of this figure we also show the projected distance of the MW pairs,  $\Delta\vec{r_{\perp}}$, illustrating that even for the first bin in $\log_{10}\Delta (J,\theta)$ these pairs are well beyond the WB regime. 
It appears that by choosing action-angle coordinates, we can trace an excess of mono-abundance stars to quite enormous distances.

It is worth mentioning that we find only a very small fraction of pairs ($\sim$ 0.2\%) having [Fe/H] $< -0.5$, and also we do not find pairs with both $[\alpha/$Fe] $> 0.15$ and [Fe/H]$< -0.5$. Therefore the contribution from thick disc stars is very small.

\section{Orbit Clustering of Stars with the Same [Fe/H]:  Friends-of-Friends Analysis} 
\label{sec:fof_algorithm}

If the ultra-wide pairs of stars with the same [Fe/H] are the descendants of dispersed birth associations, we would expect not only pairs, but triplets, quadruplets or larger associations of indistinguishable [Fe/H]. That such associations exist has been shown in the immediate vicinity of the Sun \citep{Oh2017}, using Gaia DR1. 

In Sec.~\ref{sec:pairwise_dist}, we have defined a metric (Eq.~\ref{eqn:5}) that allows us to find pairs of stars that are close in action-angle space (even at considerable distances in configuration-space). Certainly, we can use this metric to search for larger associations or structures than just pairs. One way to do this is to use the friends-of-friends (FoF), or percolation, algorithm that has been widely used in cosmology to identify features like clusters, halos or groups in density fields in N-body simulations and also in observations \citep{2010MNRAS.408.1818W,2014MNRAS.440.1763D, 2017A&C....20...44F}.

FoF algorithms can identify groups of sample members that can be linked by less than a certain  threshold distance, or ``linking length'', which can be naturally defined for the case at hand by our metric in action-angle space. If larger associations are indeed present in the data, we may expect that the contamination by chance coincidences in phase-space (occurring in our mock catalog with smooth orbit and angle distributions) should be far smaller compared to pairs, when considering ensembles of $K=3,4,...$ stars.

We will now briefly sketch the practical implementation, then show the properties of the $K>2$ associations, and their statistics in the real and mock data. It turns out that the overabundance of such associations in the real data is quite dramatic (over an order of magnitude), compared to the spurious associations in the mock data.

\subsection{Finding associations with FoF}
\label{subsec:results_fof}

To find associations based on the FoF algorithm we proceed as follows: after selecting an appropriate linking length, we limit our sample to (a) pairs that are closer than this distance, and (b) pairs that are indistinguishable in metallicity, for which we adopt
$\Delta$[Fe/H] $<0.1$. We then consecutively join all distinct pairs that have a star in common, which results in associations of $K\ge 3$ members for any linking length
$l=\log_{10} \Delta (J, \theta$); the remaining isolated pairs ($K=2$) are discarded from further consideration. In this procedure, the linking length is a free parameter, for which we will choose a range of values small enough to avoid linking vast numbers of ``field stars''. After some experimentation, we consider different linking lengths log$_{10}\,\Delta (J, \theta$):  $l_i = [-1.8,-1.7,-1.6,-1.55]$.
 
\subsection{Properties of the FoF-selected Associations}
\label{subsec:FoF_association_properties}

This FoF search yields a large number of associations with $K\ge 3$, among pairs with log$_{10}\,\Delta (J, \theta)<l_i$ that are constrained to pairwise $\Delta$[Fe/H] $<0.1$. 
We now illustrate the ensemble properties, both in action-angle space and in the space of direct observables, for a few particularly large ($K\ge 15$) candidate associations, for a linking length of log$_{10}\,\Delta (J, \theta)=-1.7$: 
the upper panel of Fig.~\ref{fig:FoF_groups} shows the distribution of these associations in proper motion, velocity-distance and position space. Among these 
nine algorithmically-identified candidate associations, six turn out to be well-known open clusters: M67, Praesepe, the Pleiades, NGC 1662, NGC 1647 and NGC 2281, labelled in the top left panel. Most of the clusters we find within this linking length, with a minimum of 15 stars per group are located at a distance between 100 and 500 pc, while M67 is located at  $\sim$ 950~pc, and NGC 1647 and NGC 2281 are located at $\sim$ 600~pc.
Their distribution in action-angle space is illustrated in the lower panel of this figure,
with the local standard of rest in these coordinates at $J_{R},J_z=0$ and $J_{\phi} \sim 1760$ 
(or $J_{\phi}=1$ in the figure).
Most of the groups show a more confined structure in action-angle space than in configuration space,
presumably by construction through the condition log$_{10}\,\Delta (J, \theta)<l_i$.  Note that the ``finite'' extent of the known clusters in action space may well result from the individual distance errors, especially for the most distant group (M67): we did not assume that the line-of-sight extent of any association should not be much larger than the transverse, angular extent (see Fig.~\ref{fig:x_y_z_HC}). Remarkably, there are also three associations with $K\ge 15$ that are just as tight in action-angle space, but widely spread in proper motion or sky position. Especially the association with black points spreads hundreds of degrees in the sky; yet it is very confined in action-angle space. The extent, distances and radial velocities of these stars seem to reveal that this group is Pisces Eridanus: the newly discovered stellar stream in Gaia DR2, which could be the remnant core of a tidally disrupted cluster or OB association \citep{2019A&A...622L..13M}. 
As a reference, in Table~\ref{table_ages} we present the ages and metallicities for some of the groups we find in Fig.~\ref{fig:FoF_groups}. These clusters have solar metallicity and are mostly young, except for M67. The age of the newly discovered Pisces Eridanus is still under debate. While \citet{2019A&A...622L..13M} claim that the age of this cluster is $\sim$ 1 Gyr, \citet{2019AJ....158...77C} find this structure to be only 120 Myr using TESS data.

We have not been able to identify the remaining two associations (in red and blue points) with known groups or clusters: they may well be newly found associations. Note from the bottom set of panels in Fig.~\ref{fig:FoF_groups} that part of the association marked with red symbols may be closely associated to the Pleiades. This only makes the point that parsing star groups into distinct entities has its limitations.

We notice that Pisces Eridanus and the group with blue points are nearly split at ($\theta_{R}$, $\theta_{z}$) = 0, respectively. The angles in \texttt{galpy} with the St{\"a}ckel approximation are defined such as $\theta_{R}$ = 0 at pericenter and increasing going towards apocenter and $\theta_{z}$ starts at zero at $z=0$ increasing towards positive $z_{max}$ \citep{galpy}. Therefore, the group with blue points for example is currently crossing the disc.

Additionally, all of these groups have low vertical action ($J_{z} < 9$ kpc km/s, Fig.~\ref{fig:FoF_groups}), and thus the harmonic oscillator approximation applies. In this regime, the frequencies are independent of the amplitudes. Consequently, our estimates for $(J_{z}, \theta_{z})$ are not strongly affected by our choice of the Galactic gravitational potential (MWPotential2014).

As we will show below, for any linking length, the number of associations grows rapidly with decreasing membership $K$. And the set of resulting associations depends of course both qualitatively (is an association found) and quantitatively (e.g. how many pairs are linked to, say, the Pleiades) on the choice of linking length. 

For the moment, we just note that our FoF approach with this  GDR2$\otimes$LMDR5 sample not only recovers algorithmically know clusters as \textquoteleft action-angle associations of indistinguishable [Fe/H]\textquoteright~and finds new ones, but also finds dispersed clusters.

Clearly, extensive follow-up of these associations is warranted.

\begin{table}
\centering
\caption{Ages and metallicities of clusters in Fig.~\ref{fig:FoF_groups}.\\
\textit{a,b}: \citet{2008A&A...484..609Y,2006A&A...450..557R}\\
\textit{c}: \citet{2018ApJ...863...67G}\\
\textit{d,e}: \citet{1997AJ....114.2556T,2015MNRAS.450.4301R}\\
\textit{f}: \citet{2013A&A...558A..53K}\\
\textit{g}: \citet{2019A&A...622L..13M,2019AJ....158...77C}}
\begin{tabular}{c|c|c|c|}
\hline
Cluster & Age (Gyr) & [Fe/H] (dex) & Ref.\\ \hline
M67 & 3.5 -- 4.8 & 0.03 $\pm$ 0.01 & \textit{a,b} \\
Praesepe & 0.65 $\pm$ 0.70 & 0.12 $\pm$ 0.04 & \textit{c}\\
Pleiades & 0.013 $\pm$ 0.005 & 0.03 $\pm$ 0.05 & \textit{c} \\
NGC 1662 & 0.42 & -0.09 & \textit{d,e}\\ 
NGC 2281 & 0.609 $\pm$ 0.013 & 0.13 $\pm$ 0.11 & \textit{f}\\
Pisces Eridanus & 0.12 -- 1 & -0.04 $\pm$ 0.15 & \textit{g}\\
\hline
\end{tabular}
\label{table_ages}
\end{table}

\subsection{Statistics of the FoF-selected Associations}
\label{subsec:FoF_association_statistics}

We now consider the basic statistics of the associations that our FoF approach identifies. If the true action-angle distribution indeed has a clustered component, while the smooth mock catalog has not, we can expect that the contrast between the real and mock data is larger for groups than for pairs alone: if \textquoteleft chance-pairs\textquoteright, drawn from a smooth orbit distribution at a given [Fe/H]
are an important contaminant, then \textquoteleft chance-triplets\textquoteright, etc. should be less so.

\begin{figure*}
\centering
\hspace*{-0.2cm}\includegraphics[width=.95\textwidth]{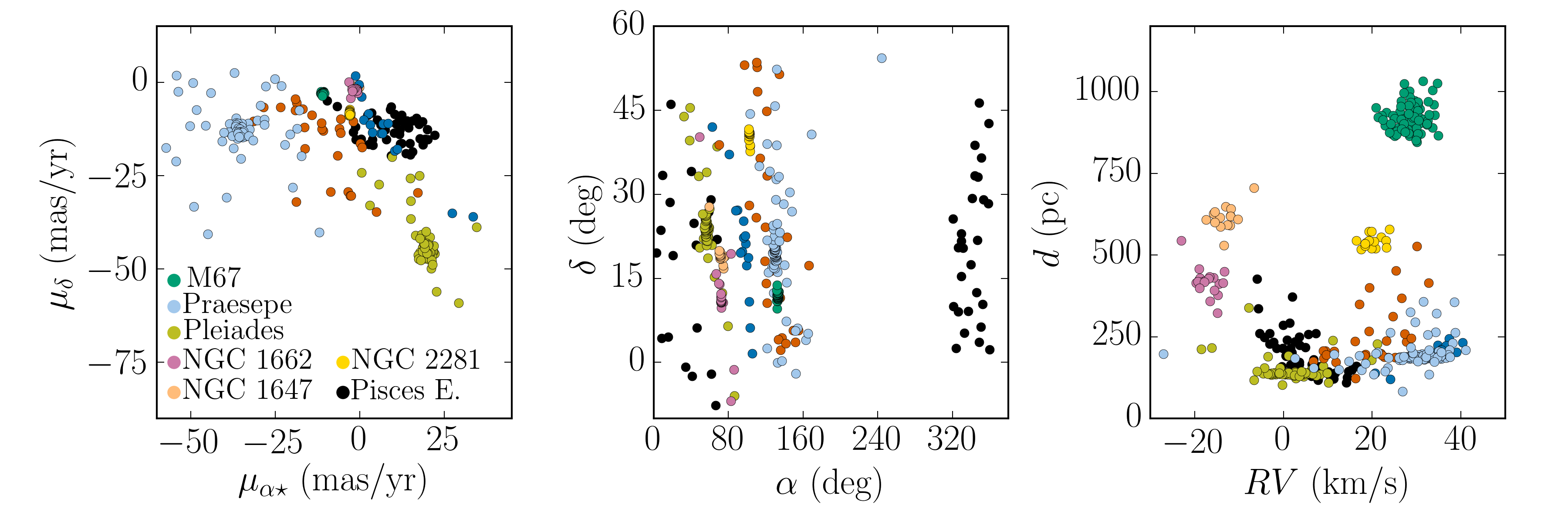}\hspace*{-0.2cm}\\
\hspace*{-0.2cm}\includegraphics[width=.95\textwidth]{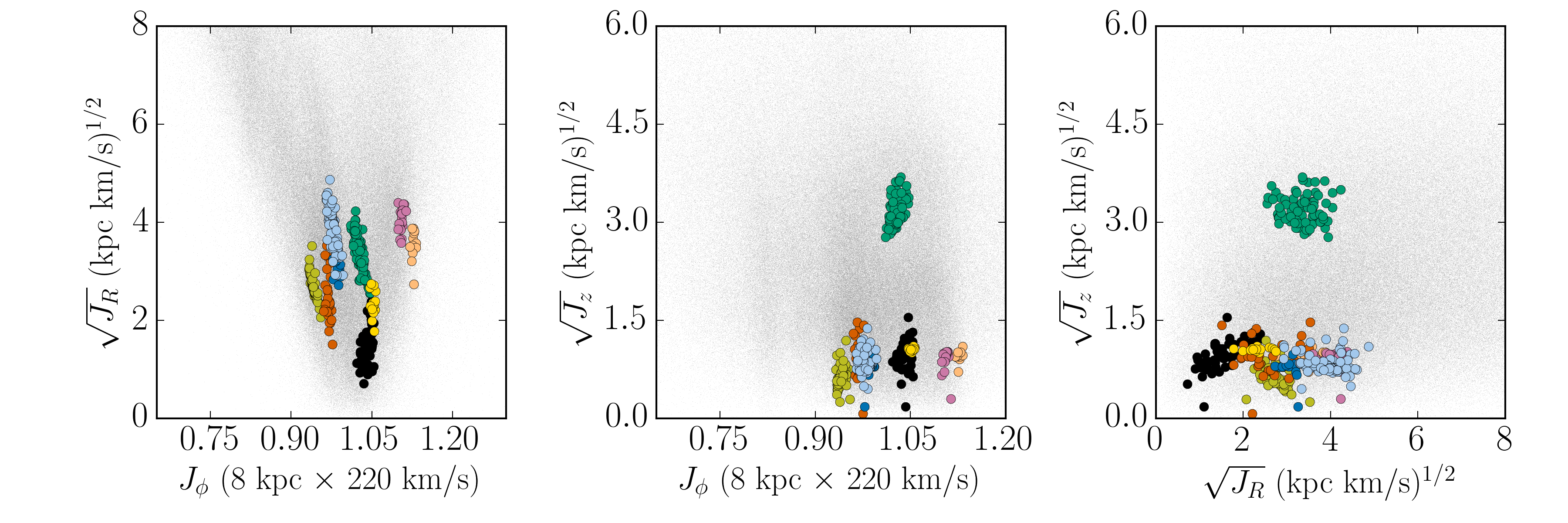}\hspace*{-0.2cm}\\
\hspace*{-0.2cm}\includegraphics[width=.95\textwidth]{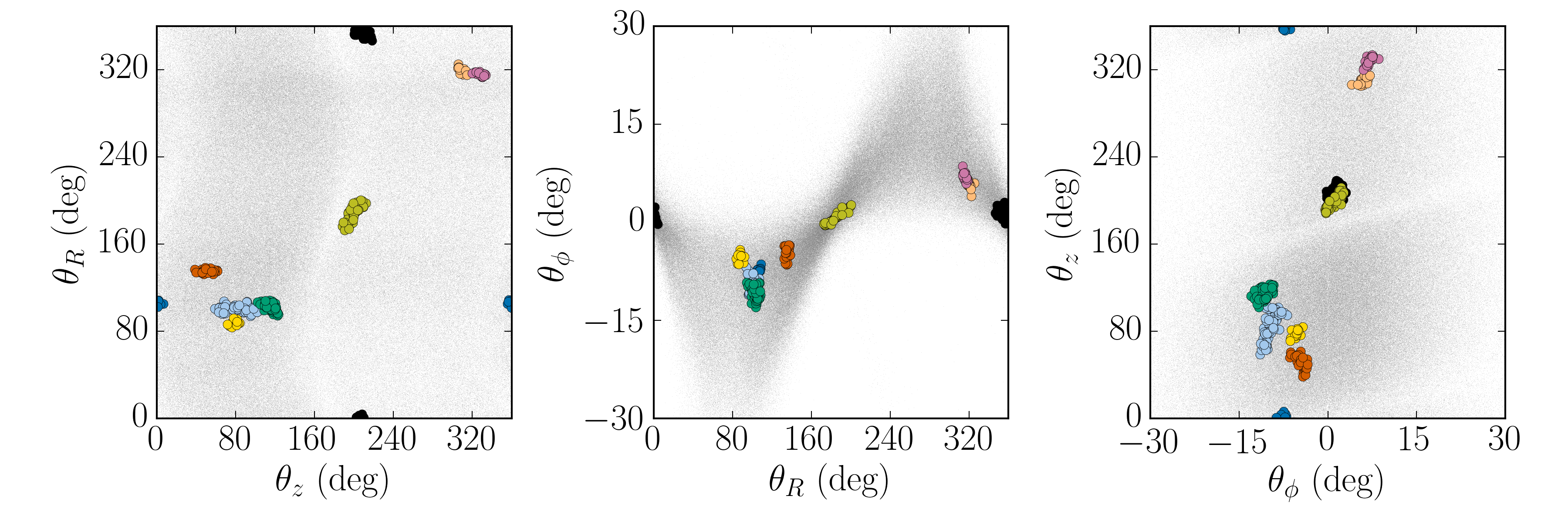}\hspace*{-0.2cm}\\

\caption{Result of our implementation of the friends-of-friends algorithm for a linking length of log$_{10}\Delta (J,\theta)$ = -1.7. We only show the largest groups we find for this specific linking length: 9 groups with a minimum of 15 members per group.\newline
\textit{Upper panels}: We show the proper motion distribution of the groups to the left, in the middle we show their position in the sky in \textit{(ra,dec)} coordinates and to the right we show their velocities and distance. We recover 7 known associations: the open cluster M67, Praesepe, the Pleiades, Pisces Eridanus, NGC 1662, NGC 1647 and NGC 2281. M67, NGC 2281, NGC 1662 and NGC 1647 (hidden behind NGC 1662) appear as concentrated clusters in $\mu_{\alpha}$ and $\mu_{\delta}$ while Praesepe and the Pleiades have a well defined center and then their structure extends further out. Most of the clusters appear as large extensions in the sky in \textit{(ra,dec)} and in velocity-distance space some of the clusters extend up to several parsec in distance. \newline
\textit{Middle and lower panels:} Here we show the three actions $J_{z}, J_{R}, J_{\phi}$ and their three respective angles, $\theta_{R}$, $\theta_{\phi}$ and $\theta_{z}$ for all the groups we find in this specific linking length. The grey dots in the background correspond to the complete dataset.
As expected by construction, all of the groups appear clustered in action and angle space. Most of the groups shown here are confined to $J_{z}<3$ (kpc km/s), only M67 reaches up to $\sim$ 9 (kpc km/s) and extends up to 12 (kpc km/s). In the radial action $J_{R}$ none of the found clusters extend beyond 25 (kpc km/s) and they are tightly constrained in $J_{\phi}$. As a reference, a star near the solar position would be located at  $J_{\phi}=1$ in this figure. From all of these groups, the black cluster (i.e., Pisces Eridanus) is the most intriguing, being very constrained in action space, but having members completely spread in \textit{ra}, separated by 240 deg in the sky. In angle space the associations are also very confined, with Pisces Eridanus having members located at $\theta_{R} \sim$ 0 and $\theta_{R} \sim$ 360 deg, showing the periodicity of the angles. The same behaviour is observed for the dark blue group but in $\theta_{z}$.}
\label{fig:FoF_groups}
\end{figure*}

\begin{figure*}
\begin{center}
\includegraphics[width=0.85\textwidth]{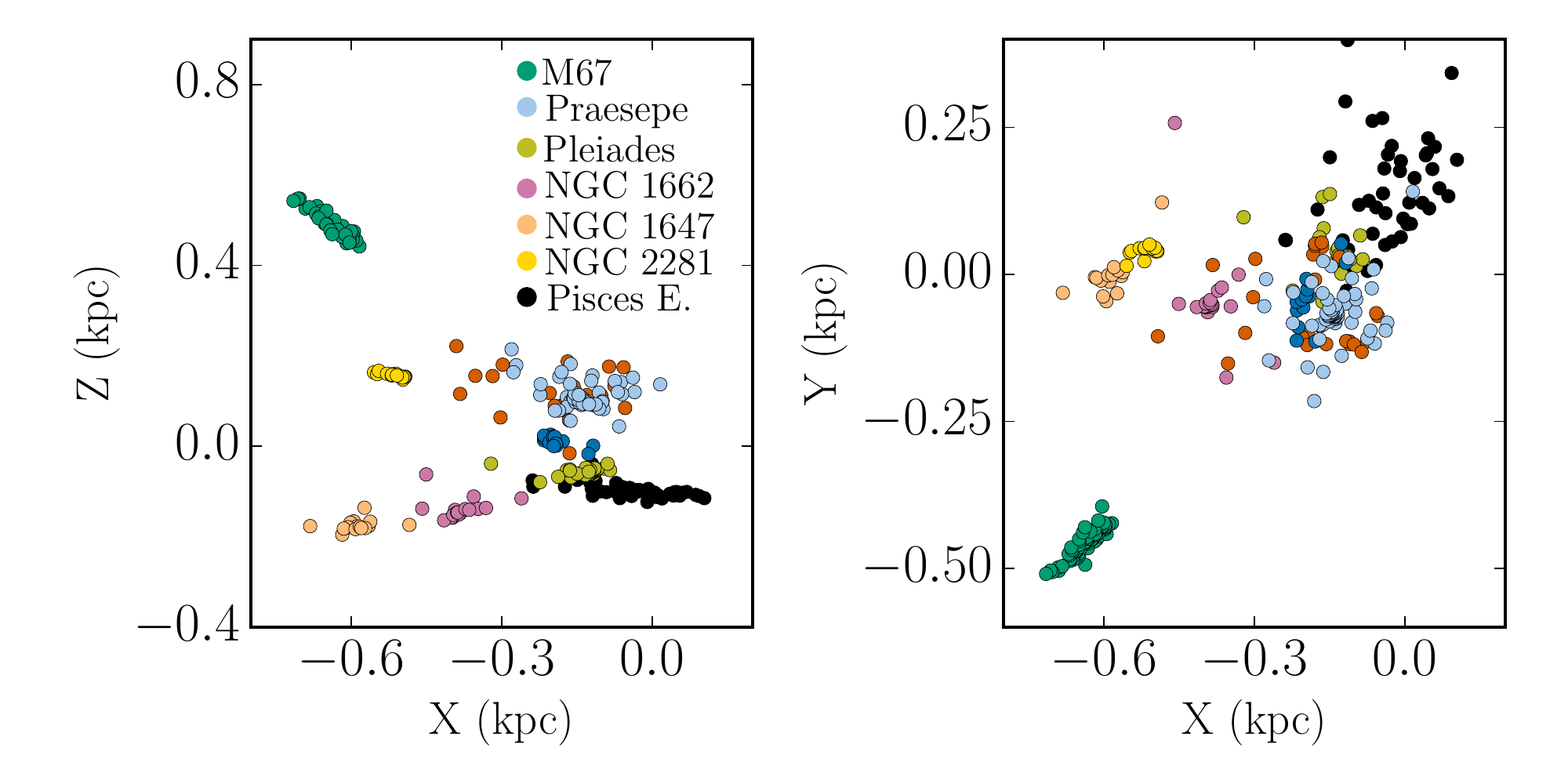}
\caption{Position in rectangular Galactic coordinates X,Y,Z of the associations we find with the FoF algorithm. These are the same as the ones presented in Fig.~\ref{fig:FoF_groups}. The Z coordinate is positive pointing towards the North Galactic pole, X increases in the direction of the Galactic center and the sun is located at (0,0,0). M67 and NGC 2281 appear mostly confined in the XZ and ZY plane, whereas Praesepe, Pisces Eridanus, the dark blue and NGC 1662 associations have members spread in the XY plane.} 
\label{fig:x_y_z_HC}
\end{center}
\end{figure*}

\begin{figure*}
\centering
\includegraphics[width=\textwidth]{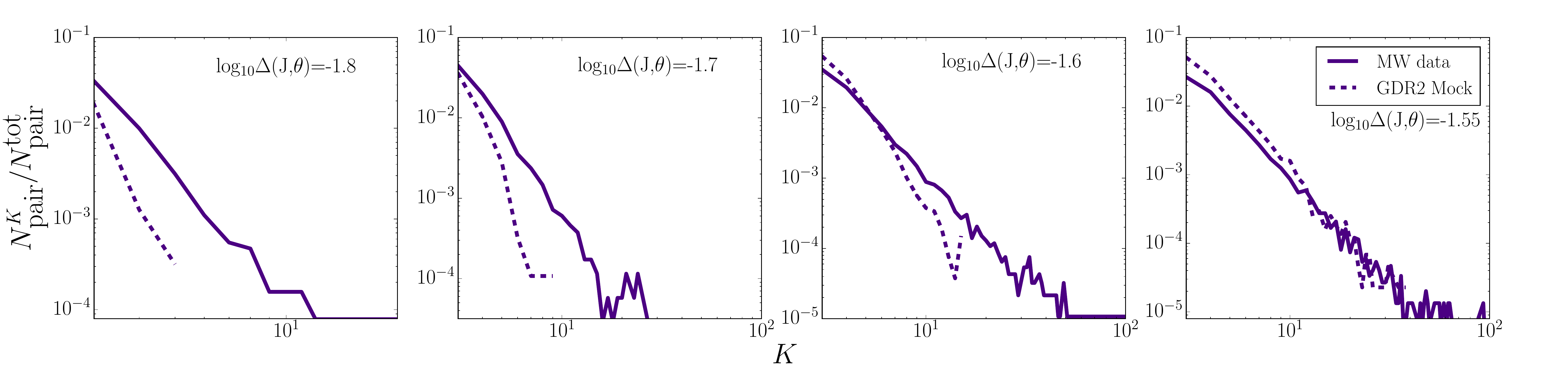}

\caption{Fraction of pairs at a given linking length that end up in FoF groups vs. the number of members in that group, $K$. The solid line reflects this fraction for the 
GDR2$\otimes$LMDR5 sample, and the dashed line to the GDR2 mock with a smooth orbit distribution. The different panels show different linking lengths, from the smallest on the left to the largest on the right. The differences between the data and the mock catalog are dramatic: associations with $K\ge 3$ are proportionally more common in the real data by an order of magnitude, except for the largest linking length. 
In the largest linking length bin (and any larger ones), the real data barely show more associations than a smooth distribution.}
\label{fig:linking_groups2}
\end{figure*}

\begin{figure*}
\centering
\includegraphics[width=\textwidth]{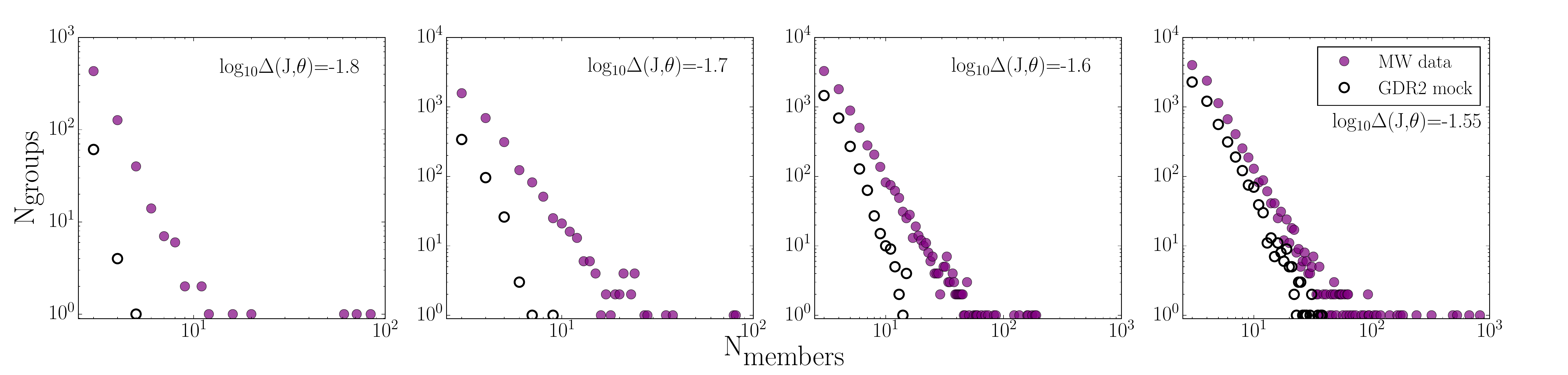}

\caption{Number of FoF groups at a given N$_{members}$, for the GDR2$\otimes$LMDR5 sample(solid circles) and GDR2 mock (open circles). The different panels show different linking lengths, from the smallest (l$_{2}$) on the left to the largest (l$_{5}$) on the right. For the shortest two linking lengths the number of FoF groups is over an order of magnitude larger, compared to GDR2 mock. At larger linking lengths these differences are less prominent. At log$_{10}\,\Delta (J, \theta) = -1.55$, the number of groups that we find for the MW data and GDR2 mock become comparable, for groups with less than 40 members; yet the largest group in the real data has $\sim 10^3$ members, the largest mock group only 38.}
\label{fig:linking_groups3}
\end{figure*}

We quantify the statistics by asking what fraction of all pairs are involved in association of ultimate size $K$, at a given linking length $l_i$; this is shown in Fig.~\ref{fig:linking_groups2}. This figure shows that for all linking lengths associations of at least 10 members are found; for $l_i=-1.6$ even 100 of them. The figure also shows that at very small linking length (e.g. $l_i=-1.8$) even the well-known clusters are not completely identified (see Fig.~\ref{fig:FoF_groups}), presumably because measurement errors push pair separations beyond this linking length. Most dramatic in this figure, is the large difference between the fraction of real (solid line) and mock (dashed line) pairs that are in associations of $K\ge 3$: for $l_i<-1.6$ there is a magnitude or more associations in the real data than in the smooth mock data. This shows quite dramatically the clustering of stars with the same [Fe/H] in action angle space, not just pairs but clearly larger ensembles or associations. This is seen more clearly in Fig.~\ref{fig:linking_groups3} where we show the number of FoF groups as a function of N$_{members}$ at different linking lengths. We find that the number of groups at a given N$_{members}$  is always at least one order of magnitude larger in the real data, compared to the mock data (where they are ``spurious'', by construction). For the largest linking length $l_i=-1.55$ the number of groups found in the mock become comparable to the real data. However, at that linking length, the largest group in the mock catalog has only N$_{members}=38$, compared to 10$^{3}$ in the real data.     

\section{Comparison to star pairs in ($\vec{r}, \vec{v}$) configuration space}

We have presented a method to calculate pairwise distances in action-angle space between MS stars in the GDR2$\otimes$LMDR5 cross-match, and we have used it to quantify the level orbit-space and abundance space clustering of the stellar distribution in our Milky Way. We have found an excess of pairs---i.e. \emph{clumping} of stars---at small phase-space distances $\Delta (J, \theta$) and small abundance differences $\Delta$[Fe/H] when compared to a mock catalog that has a smooth and phased-mixed orbital distribution (Fig.~\ref{fig:dist_action_angles_mock}).
In addition, we could show that extensive sets of star associations can be found by their orbit similarity, if they have indistinguishable [Fe/H]; we implemented the identification of these associations by a friend-of-friends algorithm.
We now provide and discuss some context for these findings. This is of course not the first time, that orbit-[Fe/H] clustering has been explored with contemporary data sets.
The recent studies by  \citet{Oh2017} and in particular by \citet{2019ApJ...884L..42K} have shown that co-moving pairs, identified by their physical separation and velocity difference, were most likely born together, as these pairs showed a strong preference for having similar metallicities. \citet{2019ApJ...884L..42K} defined a primary metric in metallicity difference $|\Delta$[Fe/H]| to determine if a pair is co-natal, and they also include the velocity and position differences of these pairs $\Delta r$ and $\Delta v$.
However, the work by \citet{2019ApJ...884L..42K} focused on pairs that are close in ($\vec{r}, \vec{v}$) with 2$<\Delta r<20$ pc and with $\Delta v<$ 1.5 km/s, and the present work -- in part by choosing action-angle coordinates --  extends to far greater distances, as we illustrate in Fig.~\ref{fig:vels_x_mock}.
We take the same bins as presented in Fig.~\ref{fig:dist_action_angles_mock2}, i.e., bins of 0.2 in log$_{10}\Delta (J,\theta)$, but now we map them into position and velocity space to illustrate to what ``typical'' distances in configuration space ($\vec{r},\vec{v}$) a certain log$_{10}\Delta (J,\theta)$ corresponds to. For example log$_{10}\Delta (J,\theta)=-1.6$, corresponds to a mean $\Delta r\sim 150$ pc and $\Delta v\sim$ 3 km/s, with many pairs
encompassing considerably greater distances in configuration space.
The selection by \citet{2019ApJ...884L..42K} is shown as a grey rectangle in  Fig.~\ref{fig:vels_x_mock},
and we see that it is closer to the properties of our wide-binary reference sample than even our smallest bin in log$_{10}\Delta (J,\theta)$, or our smallest linking length.

\section{Summary}
\label{sec:summary}
We have explored and quantified the orbit-space clustering of stars in the Galactic disc, as a function of their metallicity differences. We have done this by defining the orbit similarity between pairs of stars as the normalised distance in action-angle space $\Delta(J,\theta$), and their abundance similarity as $\Delta$[Fe/H]; we then considered both $p(\mbox{[Fe/H]}~|~\Delta(J,\theta))$ and $p(\Delta(J,\theta)~|~\mbox{[Fe/H]})$. We expect the fraction of \textquoteleft mono-abundance\textquoteright~pairs (with the same [Fe/H]) to be large for very small differences in actions {\it and} angles, $\Delta(J,\theta)$, as those stars are either wide binaries or stem from the same birth association. The fraction of mono-abundance pairs should then decrease towards larger $\Delta(J,\theta)$, as more of these star pairs on very different orbits were born at different times or at different radii, and hence have different metallicities. 

We determined the pairwise $\Delta(J,\theta$) and $\Delta$[Fe/H] for a sample of over half a million main sequence stars, with radial velocities and [Fe/H] from LAMOST and astrometric information from Gaia. Among these $\ge 10^{10}$, we found an excess of mono-abundance pairs($\Delta\mathrm{[Fe/H]}<0.1$), extending to remarkably large separations. In configuration space $(\vec{r},\vec{v})$ this $\Delta(J,\theta)$-selected excess of mono-abundance pairs extends to $\Delta r$ $\sim$ 300 pc; this is an order-of-magnitude larger than the 25~pc to which \citet{2019ApJ...884L..42K} traced it with a configuration-space selection. We assess that this is a significant \textquoteleft excess\textquoteright~through comparison with a mock sample, drawn from a smooth and phase-mixed orbit distribution with a similar selection function \citep{2018PASP..130g4101R}; in that smooth models such pairs just reflect chance similarities in action-angle space and in $\Delta$[Fe/H] (given the modest metallicity dispersion of the disc). 

We then use these action-angle distances as an input for a friends-of-friends (FoF) algorithm, to investigate which fraction of these mono-abundance pairs can be linked into larger groups (at a given linking length). Through this FoF approach, we recover a number of known clusters and associations: e.g. M67, Praesepe, the Pleaides, NGC 1662, NGC 1647 and NGC 2281. Whereas Praesepe and the Pleiades show a more extended structure in proper motion, position and distance-velocity space, the remaining known clusters are mostly confined in position and velocity space.

However, through this orbit-space FoF approach, we also find hundreds of mono-abundance associations with a very extended distribution in configuration space: extending hundreds of parsecs, and covering many degrees in the sky. For instance, we found the Pisces Eridanus stream which shows that our algorithm recovers not only clusters, but also these extended structures.

Nevertheless these stars are on similar orbits and share the same chemical information, [Fe/H]. Many of these  would not have been selected as associations in a different coordinate system.

Our analysis shows that the orbit distribution of Galactic disc stars reveals distinct small-scale clustering, among stars with indistinguishable metallicities, extending across distances of hundreds of parsec. At least qualitatively, this clustering has an obvious explanation: stars born in the same cluster, association, or even spiral-arm piece, will be born with the same [Fe/H]. Most of these birth associations will gradually disperse, as many of them may never have been gravitationally bound systems. This dispersal is driven both by orbit or action changes, which can be driven by cluster dynamics or radial migration \citep{2002MNRAS.336..785S,2018ApJ...865...96F, 2019ApJ...884L..42K}, and by the resulting orbit-phase mixing. All these effects plausibly reflect the transition from clustered star-formation to field stars. 

The results presented here suggest follow-up in various directions. On the one hand, one can use this information to quantify how effective orbit migration is in the Galactic disc. On the other hand, the work here has provided a large number of stellar association candidates. While undoubtedly some will be spurious, our FoF analysis should open a path to studying many groups of stars barely remembering their common birth origin. Follow-up could include the orbit-space search for more members in the Gaia 6D data \citep{2019MNRAS.484.3291T}, where at present metallicities are still missing. And, the LAMOST data will allow us to explore whether these associations are truly mono-abundance populations (not just of the same [Fe/H]), by looking at the other 5-10 abundances that LAMOST provides \citep{2019ApJS..245...34X}.

Overall, this analysis suggests that we may now be in a position to study the transition from clustered stars formation to field stars in an unprecedented way.

\section*{Acknowledgements}
We thank the anonymous referee for their helpful comments that improved the quality of this manuscript. Additionally, we thank Eleonora Zari, Harshil Kamdar and Charlie Conroy for useful discussions. This project was developed in part at the 2019 Santa Barbara Gaia Sprint, hosted by the Kavli Institute for Theoretical Physics at the University of California, Santa Barbara. J.C. acknowledges support from the SFB 881 program (A3) and the International Max Planck Research School for Astronomy and Cosmic Physics at Heidelberg University (IMPRS-HD). H.W.R. received support from the European Research Council under the European Union's Seventh Framework Programme (FP 7) ERC Grant Agreement n. [321035]. JR was funded by the DLR (German space agency) via grant 50\,QG\,1403. This work has made use of data from the European Space Agency (ESA) mission Gaia\footnote{\url{http://www.cosmos.esa.int/gaia}}, processed by the Gaia Data Processing and Analysis Consortium (DPAC)\footnote{\url{http://www.cosmos.esa.int/web/gaia/dpac/consortium}}. Funding for the DPAC has been provided by national institutions, in particular the institutions participating in the Gaia Multilateral Agreement. 



\bibliographystyle{mnras}
\bibliography{example} 

\newpage
\appendix
We divide the appendix in three sections. In the first one we brifly discuss the changes we apply to the model for the spectrophotometric distances presented in \citep{2018MNRAS.481.2970C}. In the second section we discuss in more detail the WB selection, and finally in the third one we show the results we obtain for a metric in action only, p(|$\Delta$  [Fe/H]| log$_{10}\Delta J$).

\section{Spectrophotometric distances}
\label{appendix_dist}
\begin{figure}

\includegraphics[width=\columnwidth]{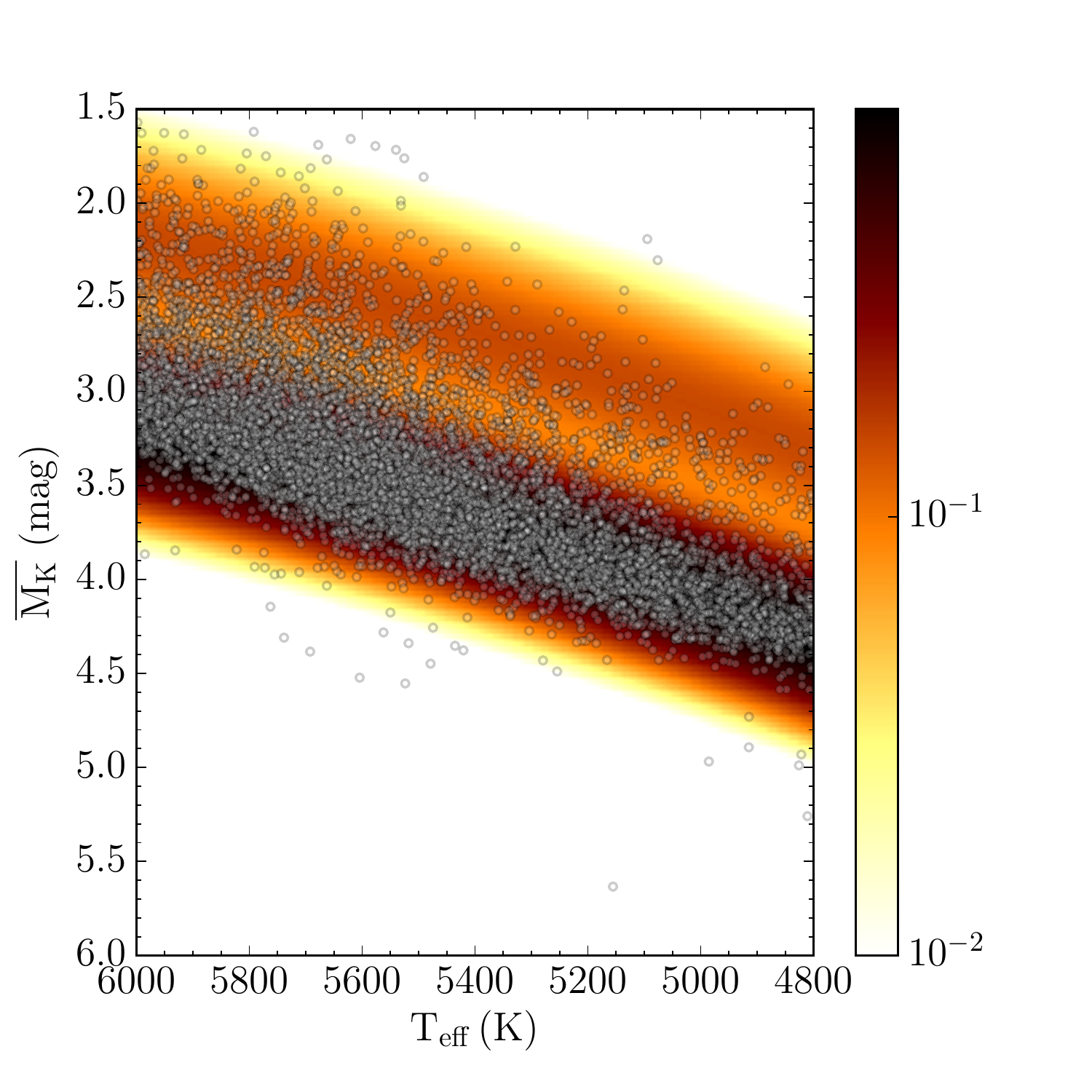}
\caption{Mean absolute magnitude model fit to MS stars in the LMDR5 sample. Importantly, the model incorporates the binary sequence. The color represents the density of the model {\it pdf} for the mean absolute magnitude. This is the best-fit model (parameters from Table~\ref{table1}), convoluted with a Gaussian of 0.15 mag reflecting the typical parallax uncertainty, for direct comparison with the data.}
\label{fig:model_pdf}
\end{figure}

In this section we show in more detail the changes that we applied to the model in \citep{2018MNRAS.481.2970C} to calculate the spectro-photometric distances with the LMDR5 $\times$ Gaia DR2 dataset used in this work. Here, we apply the same model for main sequence and binary stars defined in Sec.~3 of that work. We follow closely the same steps defined there, where the absolute magnitude of main sequence stars is a function of the spectroscopic parameters and we expand it up to first order in $\log g, \text{[Fe/H]}$ and second order in $T_\text{eff}$. However, in this case the normalization of each parameter by the mean value changes, because the dataset considers a different range in $T_\mathrm{eff}$. Hence, $\overbar{T_\mathrm{eff}}$ = 5500 K, $\overbar{\text{[Fe/H]}}$= -0.16 and $\overbar{\log g}$ = 4.4. 
In table \ref{table1} we show the new parameters obtained with $emcee$ of the best fit model for the dataset used in this work, and in Fig. \ref{fig:model_pdf} we show the mean absolute magnitude model fit to MS stars in the LMDR5 sample.

\begin{equation}
\begin{aligned}
\overbar{M_{K}}(T_\text{eff},\log g,\text{[Fe/H]} \mid \theta_K) = M_{0} + \mathrm{a_{T}}\frac{T_\mathrm{eff} - \overbar{T_\mathrm{eff}}}{\overbar{T_\mathrm{eff}}} \\
+ \mathrm{a_{T_{2}}}\left(\frac{T_\mathrm{eff} - \overbar{T_\mathrm{eff}}}{\overbar{T_\mathrm{eff}}}\right)^{2} + \mathrm{a_{\mathrm{logg}}}(\log g - \overbar{\log g})\\
+ \mathrm{a_{FeH}}\,(\mathrm{[Fe/H]} - \overbar{\mathrm{[Fe/H]}}).
\label{eq:M_K}
\end{aligned}
\end{equation}

\begin{table}
\centering
\caption{Results obtained with emcee for the parameters of our model.}
\label{table1}
\begin{tabular}{|l|l|c|}
\hline
model parameter & & best fit\\\hline
peak abs. magnitude in $K$-band & $\mathrm{M_{0}}$ (mag) & $3.602^{+0.054}_{-0.100}$\\
width of abs. magnitude distribution & $\sigma_{1}$ (mag)& $0.145 \pm 0.042$\\
prefactor of $(T_\text{eff}-\overbar{T_\text{eff}})$ term & $\mathrm{a_T}$ & $-4.935^{+0.077}_{-0.337}$\\
prefactor of $(T_\text{eff}-\overbar{T_\text{eff}})^2$ term & $\mathrm{a_{T_{2}}}$ &  $5.796^{+0.038}_{-0.150}$  \\
prefactor of $(\log g - \overbar{\log g})$ term & $\mathrm{a_{logg}}$& $1.009^{+0.132}_{-0.067}$\\
prefactor of $(\text{[Fe/H]}-\overbar{\text{[Fe/H]}})$ term & $\mathrm{a_{FeH}}$& $-0.358^{+0.133}_{-0.033}$ \\
width of binary sequence & $\mathrm{\sigma_{2}}$ (mag)& $0.268^{+0.020}_{-0.010}$ \\
binary fraction (equal mass binaries) & $f_\text{eqb}$ & $0.152 \pm 0.001$ \\ \hline
\end{tabular}
\label{table:table1}
\end{table}

\section{Cross-check with Wide Binaries}
\label{sec:WBs_appendix}
\begin{figure*}
\centering
\includegraphics[width=\textwidth]{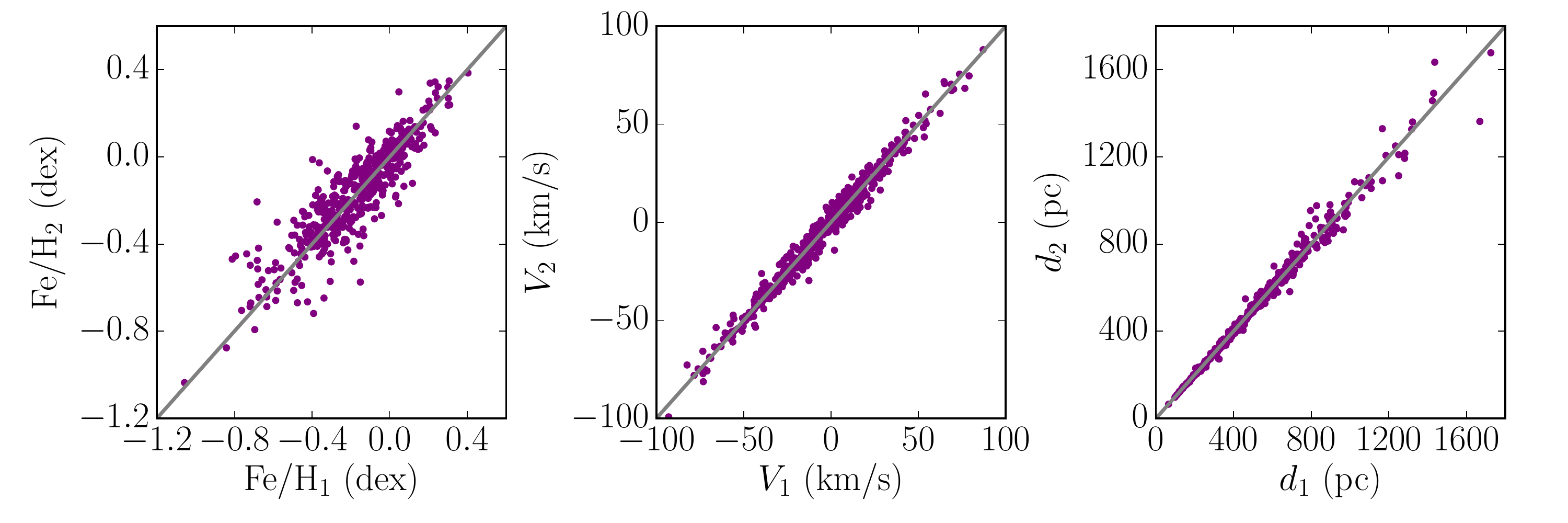}
\caption{Distribution of the WB sample in distance, velocity and metallicity, where both components show consistent velocities and distances. There is some spread in the metallicity distribution, but nonetheless, it seems mostly consistent for both components. Moreover, from Fig.~\ref{fig:hist6d} we already saw that most of the sample has differences in metallicity for these pairs of $\Delta$[Fe/H] < 0.1. In grey we overplot a 1:1 line that show how well these pairs agree. }
\label{fig:WBs}
\end{figure*}

In Fig.~\ref{fig:hist6d} we also plot (in cyan dots) the sample of WB pairs. It is expected that WBs should have similar chemical composition if they formed from the same molecular cloud \citep{2013ARA&A..51..269D}, and they are also common proper motion pairs, so their phase-space coordinates should also be consistent. Therefore, this is a good sample to compare our results to.
In Fig.~\ref{fig:WBs} we present the velocities, distances and metallicities for this sample, with all of them showing consistent values for stars in a binary system. 

We calculate the pairwise distances for each WB as defined in Eqs.~\ref{eqn:3} and \ref{eqn:4}. This distribution in Fig.~\ref{fig:hist6d} falls in the area of small log$_{10}\Delta (J,\theta)-\Delta$[Fe/H] as we would have expected: most of the WBs have log$_{10}\Delta (J,\theta)$ < -1.5 and $\Delta$[Fe/H] < 0.1 dex, with the latter corresponding to the measurement uncertainty in [Fe/H]. 
This also shows us that the features at small distances in ($J,\theta$) in the histogram of pairwise distances that we have obtained are actually real.

\section{Metric in action space only}
\label{sec:appendix_action}
\begin{figure*}
\centering
\hspace*{-0.2cm}\includegraphics[width=.45\textwidth]{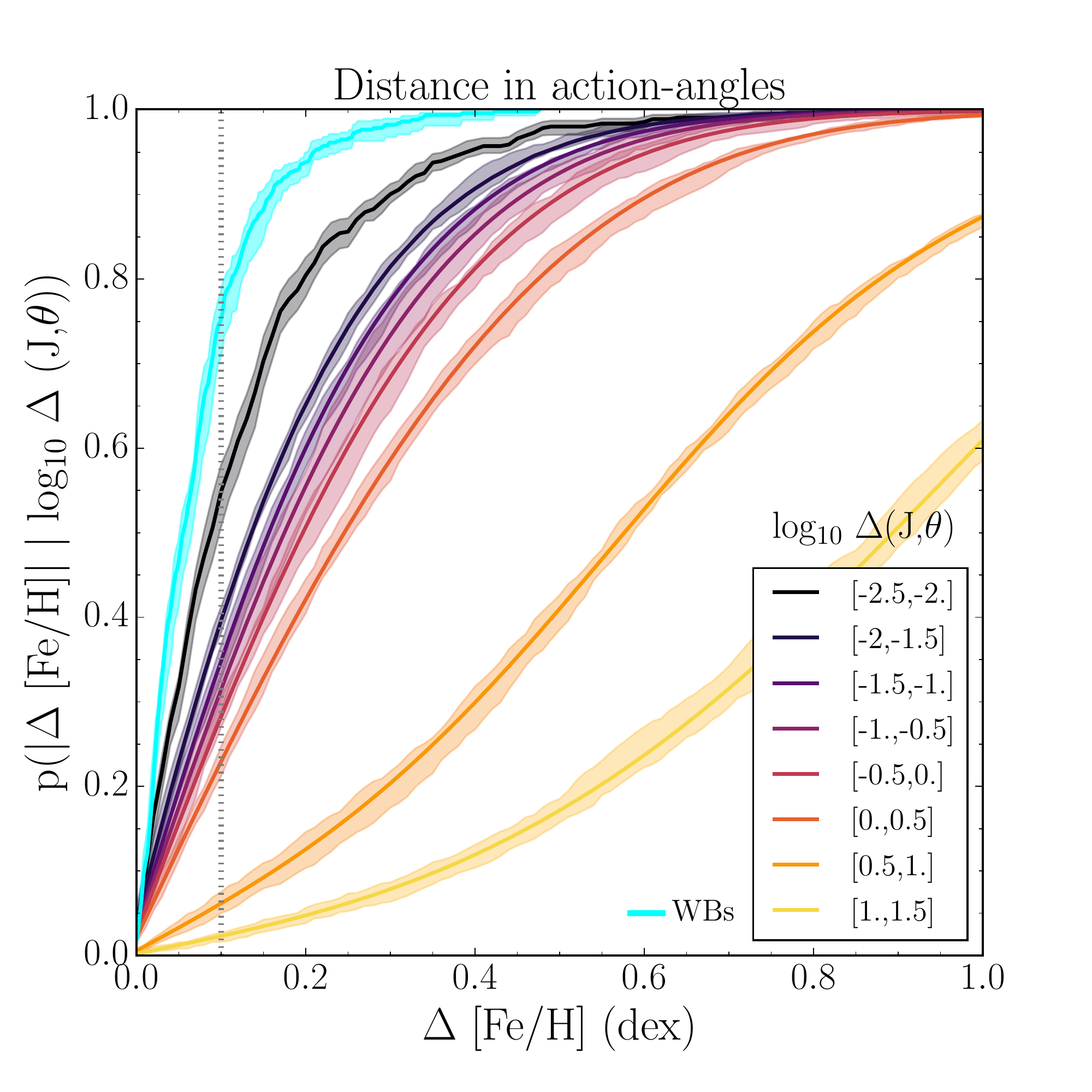}\hspace*{-0.2cm}
\hspace*{-0.2cm}\includegraphics[width=.45\textwidth]{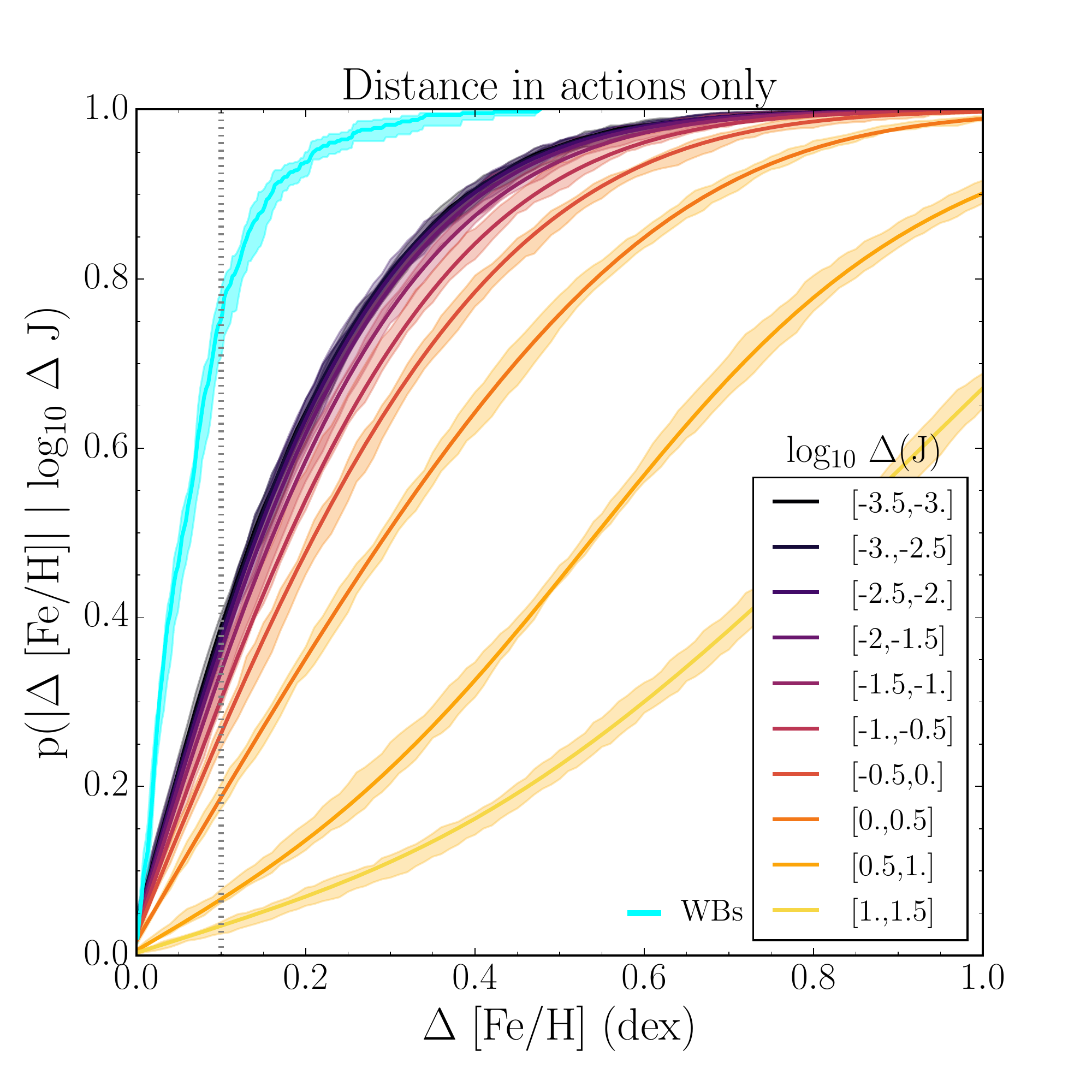}\hspace*{-0.2cm}\\
\caption{Distribution of pairwise distances in action-angle and actions only. The left side of this two panel figure is analogous to Fig.~\ref{fig:dist_action_angles_mock}. 
\newline
The right side of this figure now shows the CDF of pairs in given distance bins $\Delta (J)$ as a function of $\Delta$[Fe/H] for the LMDR5 $\times$ GDR2 MS stars. Each colored line on the left plot, again corresponds to the bins from Fig.~\ref{fig:hist6d} at different log$_{10} \Delta (J, \theta$) and to bins at different log$_{10} \Delta(J)$ on the right side.
The width of these lines show the 5th and the 95th percentile of a bootstrap re-sampling. The cyan line shows the complete distribution of WBs, for comparison. The dashed line is located at $\Delta$[Fe/H] = 0.1, that we consider as an upper limit for the uncertainties in [Fe/H].  We observe that the distance in actions only reaches smaller values than log$_{10} \Delta (J, \theta$). Even though for the first 4 bins it seems that each line lie in the same position, the rest of the bins show the same trend as in log$_{10} \Delta (J, \theta$), and we still see some signature present when considering actions only.}

\label{fig:dist_action_only}
\end{figure*}

\begin{figure*}
\centering
\hspace*{-0.2cm}\includegraphics[width=.43\textwidth]{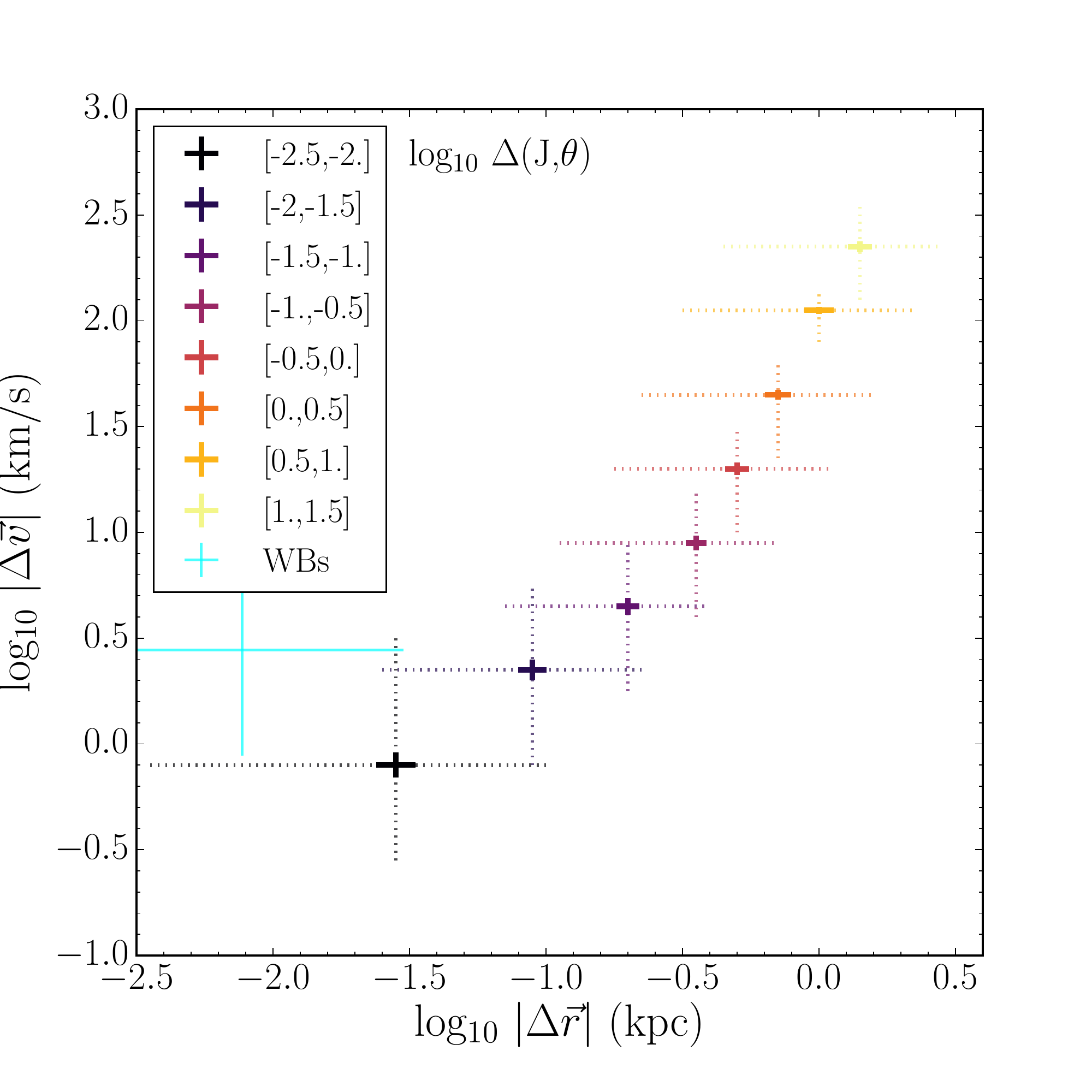}\hspace*{-0.2cm}
\hspace*{-0.2cm}\includegraphics[width=.43\textwidth]{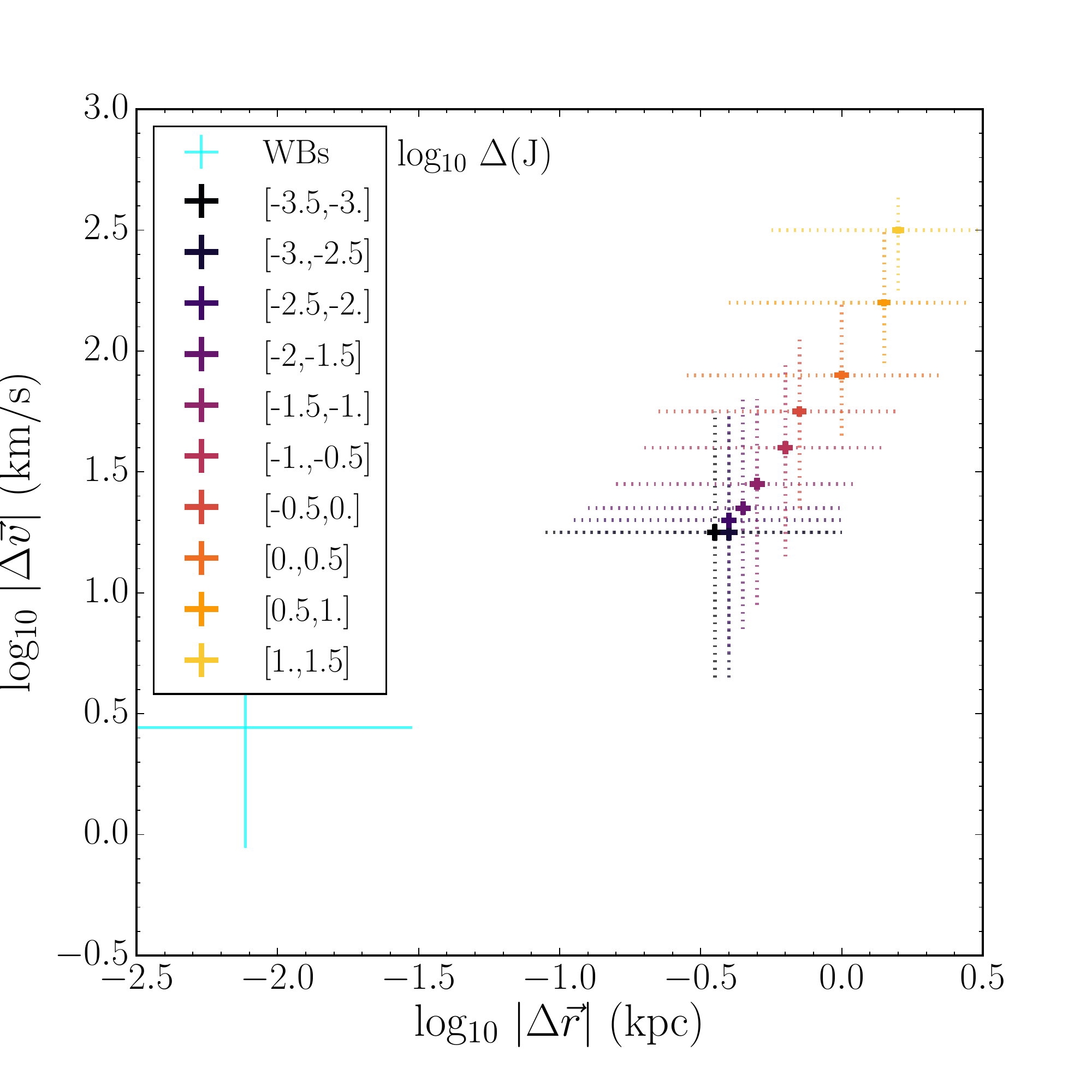}\hspace*{-0.2cm}

\caption{Differences in 3D velocities $\Delta\vec{v}$ and positions $\Delta\vec{r}$, for the same bins in log$_{10} \Delta (J, \theta$) as shown in Fig.~\ref{fig:dist_action_only} to the left, and log$_{10} \Delta(J)$ to the right. The left side of this plot is analogous to Fig.~\ref{fig:vels_x_mock}, but now we are considering bins of 0.5 in log$_{10} \Delta (J, \theta$).
 Again, we show the WBs in cyan, that are located at small $\Delta\vec{v}$ and $\Delta\vec{r}$. For the bins, the solid lines show the uncertainty of the mean value (calculated via bootstrapping), and the dashed line shows the 5th and 95th percentile. For stars close in log$_{10} \Delta(J)$ we see that it maps into large values in $\Delta\vec{r}$ and $\Delta\vec{v}$ as compared to what we observe in when we combine both actions and angles
However, this is expected as we are considering less information. We still notice that some information is present, as the first bin in log$_{10} \Delta$(J) seems to correspond to the third one in log$_{10} \Delta (J, \theta$), finding pairs with similar $\Delta (J)$-$\Delta$[Fe/H] at $\Delta(\vec{r})$=0.39 kpc, extending up to 1kpc.}
\label{fig:x_v_Jonly}
\end{figure*}

In this section we present the results of the metric in action space only. These results are not intended as a comparison to the metric in action-angle space. As we move from a 6D to 3D coordinate system, then a direct comparison is not possible. However, with these results we want to highlight that with the actions metric we still see a signature of pairs close in $\Delta J- \Delta$[Fe/H]. In the smallest bin this signal seems weaker than the one present when we include the angles, as illustrated in Fig.~\ref{fig:dist_action_only}, but again this is because we are not including the angles information. Analogous to Fig.~\ref{fig:dist_action_angles_mock}, the right side of this figure presents the results of p(|$\Delta$  [Fe/H]| log$_{10}\Delta J$). Each line here is colored at different bins of the log$_{10} \Delta J$ histogram. For the smallest bin, we find that $\sim 40\%$ of these pairs is at $\Delta$ [Fe/H] = 0.1 dex. We notice that we find smaller values of log$_{10}\Delta J$ as compared to the ones found for log$_{10}\Delta (J,\theta$), with the smallest bin at log$_{10}\Delta J = -3.5$. The first 4 bins are overlapped and don't show much difference between them, but the rest of them, from log$_{10}\Delta J = -1.5$ on wards show the same features as log$_{10}\Delta (J, \theta$). 
\newline
Finally, Fig.~\ref{fig:x_v_Jonly} shows the mapping of log$_{10} \Delta J$ into velocity-distance space to the right side. With the actions only metric we find at the smallest bin, pairs of stars between 0.1-1 kpc in $\Delta\vec{r}$ not as different to what we find with log$_{10}\Delta (J, \theta$) at 0.01-0.5 kpc. For $\Delta\vec{v}$ we find a larger difference, however the spread for log$_{10}\Delta (J, \theta$) is much larger in velocity space. 
These plots are not intended as a direct comparison between log$_{10}\Delta (J, \theta$) and log$_{10} \Delta J$. When moving from 6D to 3D coordinates inevitably we lose information. Nevertheless, we want to show that actions are still a valid coordinate system, where we can still find valuable information for pairs that are close in both log$_{10} \Delta J- \Delta$[Fe/H].  

\bsp	
\label{lastpage}
\end{document}